\documentclass[10pt,journal,compsoc]{IEEEtran}

\ifCLASSOPTIONcompsoc
  \usepackage[nocompress]{cite}
\else
  \usepackage{cite}
\fi

\ifCLASSINFOpdf
\else
\fi

\usepackage{algorithm}
\usepackage{amsmath}

\usepackage{bm}
\usepackage{array}  
\usepackage{comment}
\usepackage{threeparttable,multirow,booktabs}
\usepackage{xcolor,colortbl}
\definecolor{mygray}{gray}{0.85}

\ifCLASSINFOpdf
  \usepackage[pdftex]{graphicx}
  \graphicspath{{figs/}}
  \DeclareGraphicsExtensions{.pdf}
\else
  \usepackage[dvips]{graphicx}
  \graphicspath{{eps/}}
  \DeclareGraphicsExtensions{.eps}
\fi

\ifCLASSOPTIONcompsoc
  \usepackage[caption=false,font=normalsize,labelfont=sf,textfont=sf]{subfig}
\else
  \usepackage[caption=false,font=footnotesize]{subfig}
\fi

\usepackage{url,hyperref,bookmark}

\usepackage{pifont}
\usepackage{amssymb}

\usepackage{xspace}

\makeatletter
\DeclareRobustCommand\onedot{\futurelet\@let@token\@onedot}
\def\@onedot{\ifx\@let@token.\else.\null\fi\xspace}

\def\eg{\emph{e.g}\onedot} 
\def\ie{\emph{i.e}\onedot} 
 
\def\etc{\emph{etc}\onedot} 
 
\def\etal{\emph{et al}\onedot}
\makeatother

\newcommand{\cmark}{\ding{51}}%
\newcommand{\xmark}{\ding{55}}%

\begin{document}

\title{DAQE: Enhancing the Quality of Compressed Images by Exploiting the Inherent\\ Characteristic of Defocus}

\author{Qunliang~Xing,~\IEEEmembership{Graduate~Student~Member,~IEEE,}
        Mai~Xu,~\IEEEmembership{Senior~Member,~IEEE,}
        Xin~Deng,~\IEEEmembership{Member,~IEEE,}
        and~Yichen~Guo
\IEEEcompsocitemizethanks{\IEEEcompsocthanksitem Q. Xing is affiliated with the School of Electronic Information Engineering and the Shen Yuan Honors College, Beihang University, Beijing, China. E-mail: xingql@buaa.edu.cn.
\IEEEcompsocthanksitem M.~Xu and Y.~Guo are affiliated with the School of Electronic Information Engineering, Beihang University, Beijing, China. E-mail: \{maixu,gyc970930\}@buaa.edu.cn.
\IEEEcompsocthanksitem X. Deng is affiliated with the School of Cyber Science and Technology, Beihang University, Beijing, China. E-mail: cindydeng@buaa.edu.cn.
\IEEEcompsocthanksitem Corresponding author: Mai Xu.}
}

\markboth{DAQE: Enhancing the Quality of Compressed Images by Exploiting the Inherent Characteristic of Defocus}%
{Xing \MakeLowercase{\textit{et al.}}: DAQE: Enhancing the Quality of Compressed Images by Exploiting the Inherent Characteristic of Defocus}

\IEEEtitleabstractindextext{%
\begin{abstract}
  Image defocus is inherent in the physics of image formation caused by the optical aberration of lenses, providing plentiful information on image quality.
  Unfortunately, existing quality enhancement approaches for compressed images neglect the inherent characteristic of defocus, resulting in inferior performance.
  This paper finds that in compressed images, significantly defocused regions have better compression quality, and two regions with different defocus values possess diverse texture patterns.
  These observations motivate our defocus-aware quality enhancement (DAQE) approach.
  Specifically, we propose a novel dynamic region-based deep learning architecture of the DAQE approach, which considers the regionwise defocus difference of compressed images in two aspects.
  (1) The DAQE approach employs fewer computational resources to enhance the quality of significantly defocused regions and more resources to enhance the quality of other regions;
  (2) The DAQE approach learns to separately enhance diverse texture patterns for regions with different defocus values, such that texture-specific enhancement can be achieved.
  Extensive experiments validate the superiority of our DAQE approach over state-of-the-art approaches in terms of quality enhancement and resource savings.
\end{abstract}

}

\maketitle

\IEEEdisplaynontitleabstractindextext

\IEEEpeerreviewmaketitle

\IEEEraisesectionheading{\section{Introduction}\label{sec-introduction}}

\IEEEPARstart{N}{owadays}, we are embracing an era of the explosive growth of images.
According to Domo statistics~\cite{inc.DataNeverSleeps2020}, Facebook stored and transmitted approximately 147,000 images per minute in 2020;
similar situations were observed in other internet servers, such as WeChat and Twitter.
To store and transmit such a large number of images, several lossy image compression standards, \eg, joint photographic experts group (JPEG)~\cite{wallaceJPEGStillPicture1992}, JPEG 2000~\cite{marcellinOverviewJPEG20002000}, and high-efficiency video coding with main still image profile (HEVC-MSP)/better portable graphics (BPG)~\cite{sullivanOverviewHighEfficiency2012,bellardBetterPortableGraphics2018}, have been successfully developed to reduce transmission bandwidths and storage costs.
However, the compressed images suffer from compression artifacts, \eg, ringing, blocking, and blurring effects~\cite{shenReviewPostprocessingTechniques1998}, thus degrading the quality of user experience (QoE)~\cite{seshadrinathanStudySubjectiveObjective2010,itu-t10VocabularyPerformance2017}.

This paper proposes enhancing the quality of compressed images by taking into account the characteristic of image defocus, which is a blurring effect caused by the optical aberrations of lenses.
Specifically, only regions close to the focal plane, \ie, within the depth of field (DoF)~\cite{salvaggioBasicPhotographicMaterials2013}, appear to be focused, while regions far from the focal plane are blurred~\cite{krausDepthoffieldRenderingPyramidal2007}.
Given the characteristic of image defocus, there are two main drawbacks of state-of-the-art approaches~\cite{dongCompressionArtifactsReduction2015,guoBuildingDualdomainRepresentations2016,wangD3DeepDualdomain2016,galteriDeepGenerativeAdversarial2017,zhangGaussianDenoiserResidual2017,wangNovelDeepLearningbased2017,guoOnetomanyNetworkVisually2017,maoEnhancedImageDecoding2018,maoFidelityQualityRegionaware2019,xingEarlyExitNot2020,dengSpatioTemporalDeformableConvolution2020,guanMFQENewApproach2021,zhengProgressiveTrainingTwoStage2022,yangNTIRE2022Challenge2022} regarding the quality enhancement of compressed images.
\textbf{(1) Regionwise quality agnostic.}
Existing approaches neglect the difference in the quality of different regions of an input image;
thus, they process the whole image in the same manner.
However, there exists a significant regionwise quality difference in a single compressed image, particularly referring to regions with different defocus values.
\textbf{(2) Regionwise texture agnostic.}
Existing approaches do not consider the texture difference in a compressed image.
Consequently, they are not effective in enhancing diverse texture patterns, of which the diversity can also be reflected in their defocus values.
Ideally, texture-specific quality enhancement should be conducted for diverse texture patterns, especially those of regions with different defocus values.

\begin{figure*}[!t]
  \centering
  \includegraphics[width=.95\linewidth]{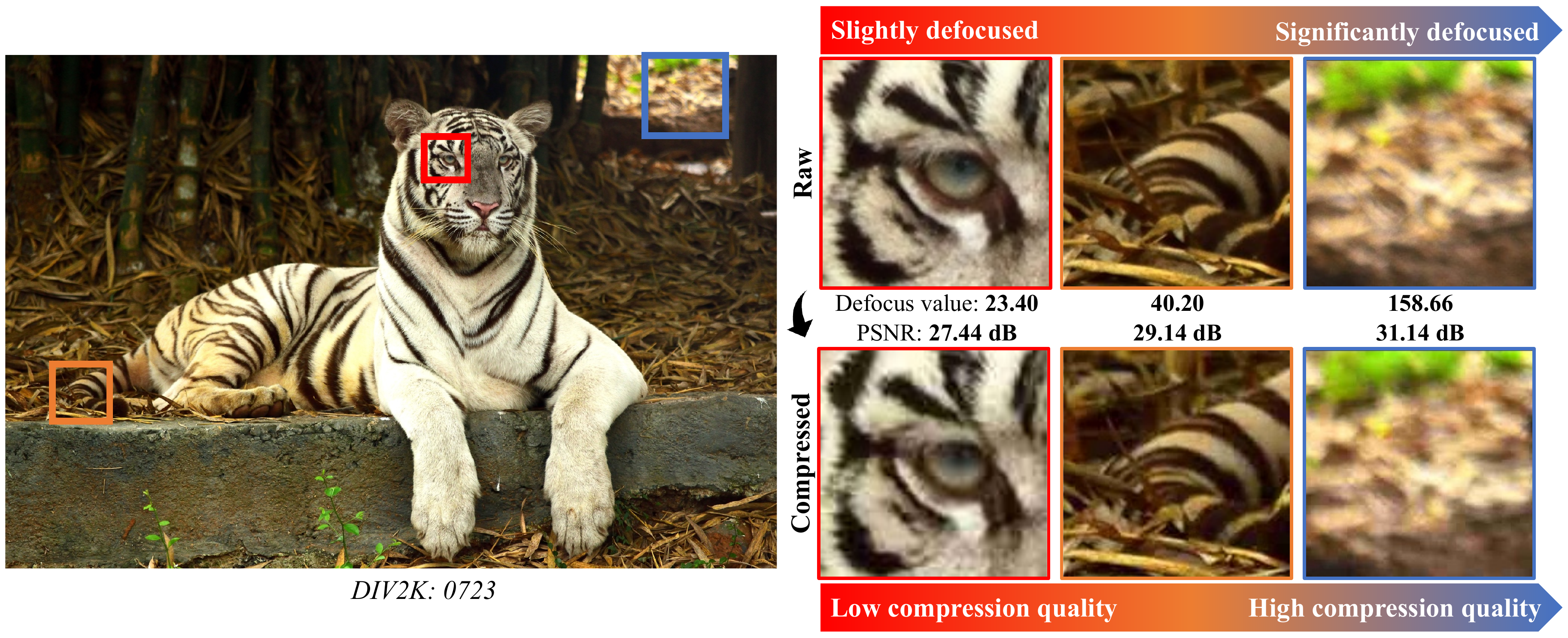}
  \caption{Motivation of our DAQE approach.
  There exist regions with different defocus values within an image.
  The defocus values are estimated by the defocus map estimation network (DMENet)~\cite{leeDeepDefocusMap2019}.
  The image is compressed by JPEG~\cite{wallaceJPEGStillPicture1992} with the quality factor (QF) as 20.}
  \label{fig-fig1}
\end{figure*}

We address the above two drawbacks of existing approaches by utilizing inherent and off-the-shelf image defocus.
We obtain two observations by analyzing the defocus and quality of compressed images from the diverse 2K resolution image (DIV2K) dataset~\cite{agustssonNTIRE2017Challenge2017}, as shown in Figure~\ref{fig-fig1}.
(1) The compression quality of compressed images is highly correlated with image defocus.
Specifically, in a compressed image, significantly defocused regions have better compression quality than slightly defocused regions.
Thus, regions with different defocus values in a compressed image should be separately enhanced.
(2)
Regions with different defocus values tend to have diverse texture patterns.
Therefore, texture-specific enhancement can be achieved by separately enhancing regions with different defocus values.

Based on our observations, we propose a defocus-aware quality enhancement approach, named DAQE, for enhancing the quality of compressed images.
The DAQE approach is equipped with a novel dynamic deep learning-based architecture.
First, the DAQE approach estimates the defocus map for the input image.
Then, the DAQE approach conducts patchwise dynamic enhancement for patches with different defocus values separately.
Considering that significantly defocused patches have superior compression quality compared with slightly defocused patches, the DAQE approach employs fewer computational resources to enhance the quality of significantly defocused patches and more resources on other patches to improve efficiency.
Note that all patches are enhanced with a single dynamic architecture in an ``easy-to-hard'' manner.
Additionally, the DAQE approach extracts diverse texture patterns for patches with different defocus values by embedding a unique attention-based texture learner in each enhancement path.
The texture learner is designed to extract texture patterns diverse in shape and intensity.
In this way, we can achieve texture-specific quality enhancement with improved efficacy.

Finally, we conduct extensive experiments to validate the effectiveness of our DAQE approach in terms of quality enhancement and resource savings, which is significantly better than state-of-the-art approaches.
Furthermore, we demonstrate the effectiveness of utilizing defocus for quality enhancement in two aspects:
(1) Explicitly clustering patches with different quality is effective for enhancing the quality of compressed images, which can be approached efficiently by using image defocus;
(2) Explicitly reasoning about defocus reduces the difficulty of finding relevance among global patches.
Our demonstrations explain the success of DAQE for quality enhancement of compressed images, and also have the potential to inspire other regionwise image enhancement works.

\section{Related Works}\label{sec-related}

\subsection{Quality Enhancement of Compressed Images}

During the past decade, many deep learning-based approaches~\cite{dongCompressionArtifactsReduction2015,guoBuildingDualdomainRepresentations2016,wangD3DeepDualdomain2016,zhangGaussianDenoiserResidual2017,wangNovelDeepLearningbased2017,xingEarlyExitNot2020} have been proposed for enhancing the quality of compressed images, owing to the successful development of convolutional neural networks (CNNs)~\cite{lecunDeepLearning2015}.
Specifically, Dong \etal~\cite{dongCompressionArtifactsReduction2015} proposed a shallow four-layer artifact reduction CNN (AR-CNN), pioneering CNN-based quality enhancement approaches for JPEG-compressed images.
Later, approaches with deeper CNN structures and the quantization prior of JPEG compression, \ie, deep dual-domain (D3)~\cite{wangD3DeepDualdomain2016} and deep dual-domain CNN (DDCN)~\cite{guoBuildingDualdomainRepresentations2016}, were proposed to remove JPEG compression artifacts.
Wang \etal~\cite{wangNovelDeepLearningbased2017} proposed a 10-layer deep CNN-based auto decoder (DCAD), which is the first CNN-based quality enhancement approach for BPG-compressed images.
DCAD does not utilize coding information from codecs but surpasses most previous approaches in terms of the quality of enhanced images thanks to the effective learning structure of a much deeper network.
To take a step forward, the denoising convolutional neural network (DnCNN)~\cite{zhangGaussianDenoiserResidual2017} was proposed, which combines a 20-layer deep network with advanced techniques of the day including residual learning~\cite{heDeepResidualLearning2016} and batch normalization~\cite{ioffeBatchNormalizationAccelerating2015}.
In this way, DnCNN significantly outperforms most traditional model-based approaches such as block-matching and 3-D filtering (BM3D)~\cite{dabovImageDenoisingSparse2007}, as well as the above learning-based approaches.
Most recently, Xing \etal~\cite{xingEarlyExitNot2020} proposed a resource-efficient blind quality enhancement (RBQE) approach for both JPEG-compressed and BPG-compressed images.
The RBQE approach was designed with a dynamic inference structure, such that blind yet effective quality enhancement can be achieved for compressed images.
In this paper, we propose utilizing image defocus for the quality enhancement of compressed images.

\begin{figure*}[!t]
  \centering
  \includegraphics[width=1.\linewidth]{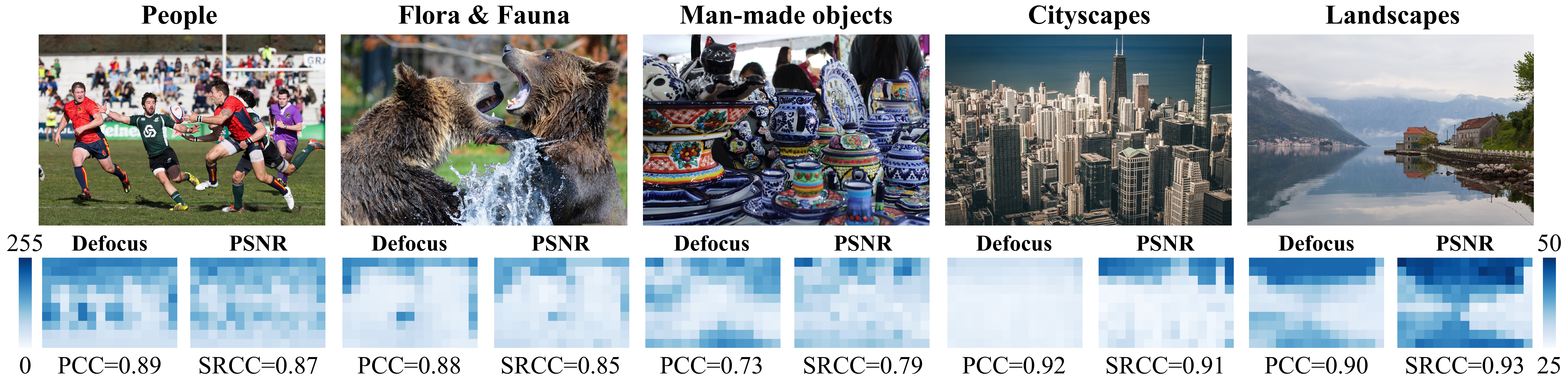}
  \caption{Correlation between the patchwise defocus and PSNR (dB) values within a single image.
  Five example images from the DIV2K dataset with different contents are presented.
  The left color bar is for defocus maps, while the right one is for PSNR maps.
  Images are compressed by the BPG codec with the quantization parameter (QP) as 37.}
  \label{fig-dataset}
\end{figure*}

\subsection{Defocus-Aware Vision Tasks}

In this section, we review defocus-aware works of related vision tasks.
The characteristic of image defocus provides plentiful information about image quality, depth, objectness, saliency, \etc.
Hence, image defocus has been widely used in many vision tasks, \eg, image depth estimation~\cite{pentlandDepthSceneDepth1982,ziouDepthDefocusEstimation2001,gurSingleImageDepth2019,maximovFocusDefocusBridging2020}, image defocus deblurring~\cite{gureyevImageDeblurringMeans2004,zhouWhatAreGood2009,zhangSpatiallyVariantDefocus2016}, image saliency detection~\cite{chenHybridSaliencyDetection2013,jiangSalientRegionDetection2013}, and image segmentation~\cite{swainDefocusbasedImageSegmentation1995}.
For image depth estimation, Pentland \etal~\cite{pentlandDepthSceneDepth1982} showed that two images formed with different apertures indicate depth information.
Thus, the image depth can be generated from image defocus.
For image defocus deblurring, the works of \cite{gureyevImageDeblurringMeans2004,zhouWhatAreGood2009,zhangSpatiallyVariantDefocus2016} are relevant, in which the defocus kernel is estimated and then used to deblur images.
For image saliency prediction, Jiang \etal~\cite{jiangSalientRegionDetection2013} found that salient image regions are often photographed in focus;
therefore, the estimation of image defocus maps can boost the performance of higher-level saliency prediction.

To the best of our knowledge, no works consider defocus-aware quality enhancement for compressed images.
In addition, the correlation between the region quality and region defocus of compressed images is unclear.
In this paper, we thoroughly investigate this correlation and demonstrate that the characteristic of image defocus can significantly benefit quality enhancement by our proposed DAQE approach.

\section{Observations}\label{sec-observations}

This section presents our observations on how the characteristic of defocus is related to the regionwise quality and texture patterns of the compressed images.
Our observations are obtained by analyzing a widely used DIV2K dataset~\cite{agustssonNTIRE2017Challenge2017}, which includes 900 images with 2K resolution.
These images cover a large diversity of contents, including people (13.67\%), flora and fauna (31.56\%), man-made objects (19.11\%), cityscapes (20.78\%), and landscapes (14.89\%), as shown in Figure~\ref{fig-dataset}.
In addition, these images can also fall into the scenes of indoor (11.89\%), outdoor (83.89\%), and underwater (4.22\%).
First, to evaluate the defocus level for each image, we adopt state-of-the-art DMENet~\cite{leeDeepDefocusMap2019} to generate a defocus map for each image.\footnote{A defocus map is an eight-bit grayscale image ranging from 0 to 255. Pixels with larger defocus values are estimated to be further away from the focal plane.}
Then, to obtain compressed images, we compress all images with two compression codecs (\ie, the BPG~\cite{bellardBetterPortableGraphics2018} and JPEG~\cite{wallaceJPEGStillPicture1992} codecs) and eight settings (\ie, with a quantization parameter (QP)~\cite{sullivanOverviewHighEfficiency2012} of 27/32/37/42 or a quality factor (QF) of 20/30/40/50).
Next, to evaluate the regionwise defocus, quality, and texture patterns, we crop all images and defocus maps into nonoverlapping patches with size $128 \times 128$.
Finally, we calculate the average defocus value for each patch as the patchwise defocus value.

\textbf{Observation 1:}
There exists dramatic variation in the patchwise defocus values within a single image.

\begin{table}[!t]
  \caption{Variation in the patchwise defocus values within a single image.}
  \label{tab-defocus-difference}
  \centering
  \begin{tabular}{l | cccc}
    \toprule
    Content & People & Flora\&Fauna & Man-made & Cityscapes \\
    \midrule
    STD & 34.26 & 39.13 & 26.06 & 37.79 \\
    Mean  & 64.64 & 69.99 & 50.00 & 55.17 \\
    CV (\%) & 50.16 & 54.74 & 46.67 & 61.10 \\
    \midrule
    Range & 129.16 & 135.84 & 107.26 & 130.67 \\
    \bottomrule
    \toprule
    Content & Landscapes & Indoor & Outdoor & Underwater \\
    \midrule
    STD & 34.81 & 24.85 & 36.67 & 31.44 \\
    Mean & 52.99 & 52.99 & 60.91 & 57.58 \\
    CV (\%) & 60.66 & 41.57 & 56.82 & 51.34 \\
    \midrule
    Range & 127.07 & 103.36 & 130.69 & 122.24 \\
    \bottomrule
  \end{tabular}
\end{table}

\begin{figure}[!t]
  \centering
  \includegraphics[width=1.\linewidth]{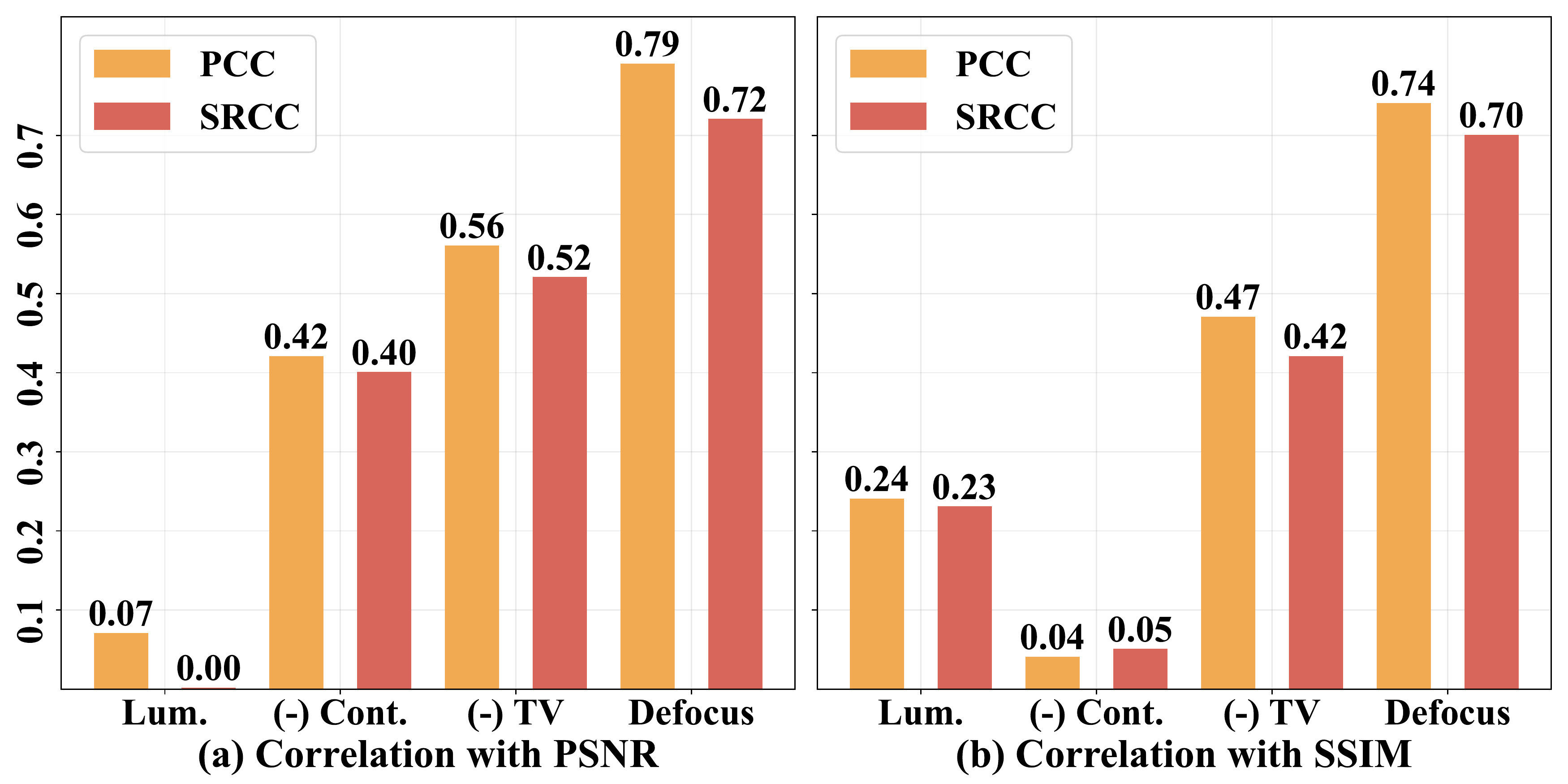}
  \caption{Correlation between patch quality and features.
  The ``lum'' and ``cont'' are the abbreviations of ``luminance'' and ``contrast''.}
  \label{fig-defocus-quality}
\end{figure}

\textbf{Analysis:}
We measure the variation in the patchwise defocus values in a single image in terms of the standard deviation (STD), coefficient of variation (CV)~\cite{everittCambridgeDictionaryStatistics2011}, and range.
Specifically, the CV value is the ratio of the STD value to the mean value.
The range value is obtained by subtracting the lowest patchwise defocus value from the highest value within an image.
As shown in Table~\ref{tab-defocus-difference}, the CV value is no less than 40\% for all contents, indicating a strong variation in patchwise defocus values.
In addition, the defocus range is up to 135.84 (\ie, for Flora \& Fauna), which is nearly two times the corresponding mean value (\ie, 69.99).
Similar results can be found for other contents in Table~\ref{tab-defocus-difference}, implying a large interval of the patchwise defocus values within each image.
Thus, the variation in patchwise defocus values within a single image is dramatic.
Intuitively, a shallow DoF is typically preferred by photographers to produce high-quality images, causing the difference in patchwise defocus values.
Hence, the widely-used DIV2K benchmark for quality enhancement can exhibit such a large variation in the patchwise defocus values.
Finally, the analysis of Observation 1 is accomplished.

\textbf{Observation 2:}
For a compressed image, patches with higher defocus values tend to have better compression quality.

\textbf{Analysis:}
We adopt two widely-used quality assessment metrics, \ie, the peak signal-to-noise ratio (PSNR) and structural similarity index measure (SSIM)~\cite{wangImageQualityAssessment2004}, for measuring the compression quality.
Then, for each compression setting, the Pearson correlation coefficient (PCC)~\cite{pearsonVIINoteRegression1895} and Spearman's rank correlation coefficient (SRCC)~\cite{spearmanProofMeasurementAssociation1904} values are calculated between the defocus and quality values for all patches, to validate their correlation.
The results are then averaged by eight compression settings.
In addition to the defocus, we adopt the features of luminance, contrast, and total variation (TV) as the baseline.
As shown in Figure~\ref{fig-defocus-quality}, both the PSNR and SSIM values of patches are highly correlated with their corresponding defocus values.
Specifically, both the PCC and SRCC values between quality and defocus are above 0.70, significantly higher than those between quality and baseline features.
Some examples are shown in Figure~\ref{fig-dataset}.
Consequently, defocus can serve as a good indicator for regionwise compression quality (measured by PSNR and SSIM).
More importantly, the correlation between defocus and quality is positive, implying superior compression quality for patches with higher defocus values.
In fact, for better rate-distortion performance, image compression is mainly performed on high-frequency components~\cite{wallaceJPEGStillPicture1992,sullivanOverviewHighEfficiency2012};
as a result, patches with higher defocus values tend to have better compression quality due to the weakened high-frequency components caused by defocus.
Finally, the analysis of Observation 2 is accomplished.

\textbf{Observation 3:}
For a compressed image, the texture patterns of the patches with dissimilar defocus values are more diverse than those with similar defocus values.

\begin{figure}[!t]
  \centering
  \includegraphics[width=1.\linewidth]{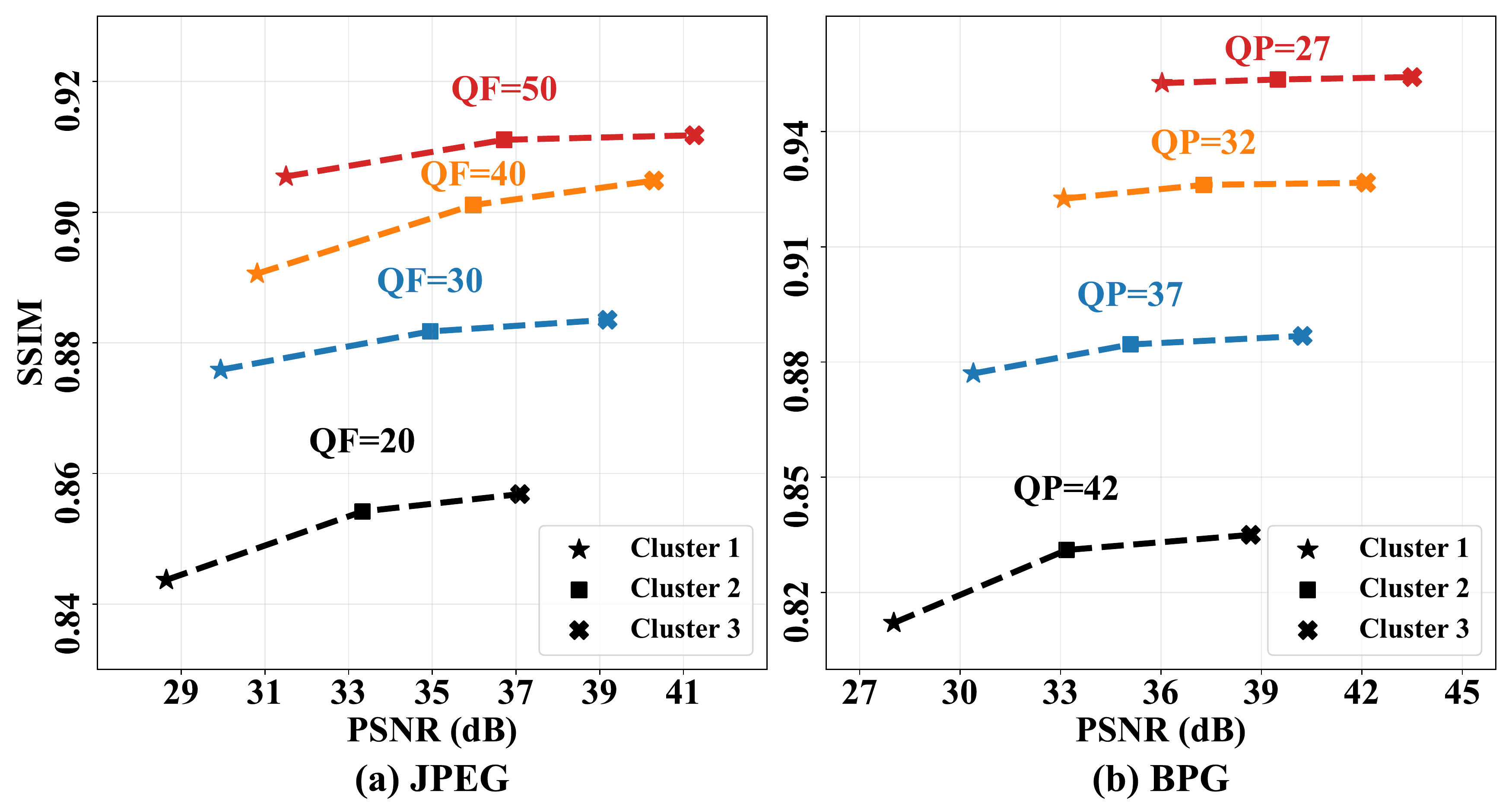}
  \caption{Average patch quality of three clusters in terms of PSNR (dB) and SSIM.}
  \label{fig-observations-defocus-cluster}
\end{figure}

\begin{figure}[!t]
  \centering
  \includegraphics[width=1.\linewidth]{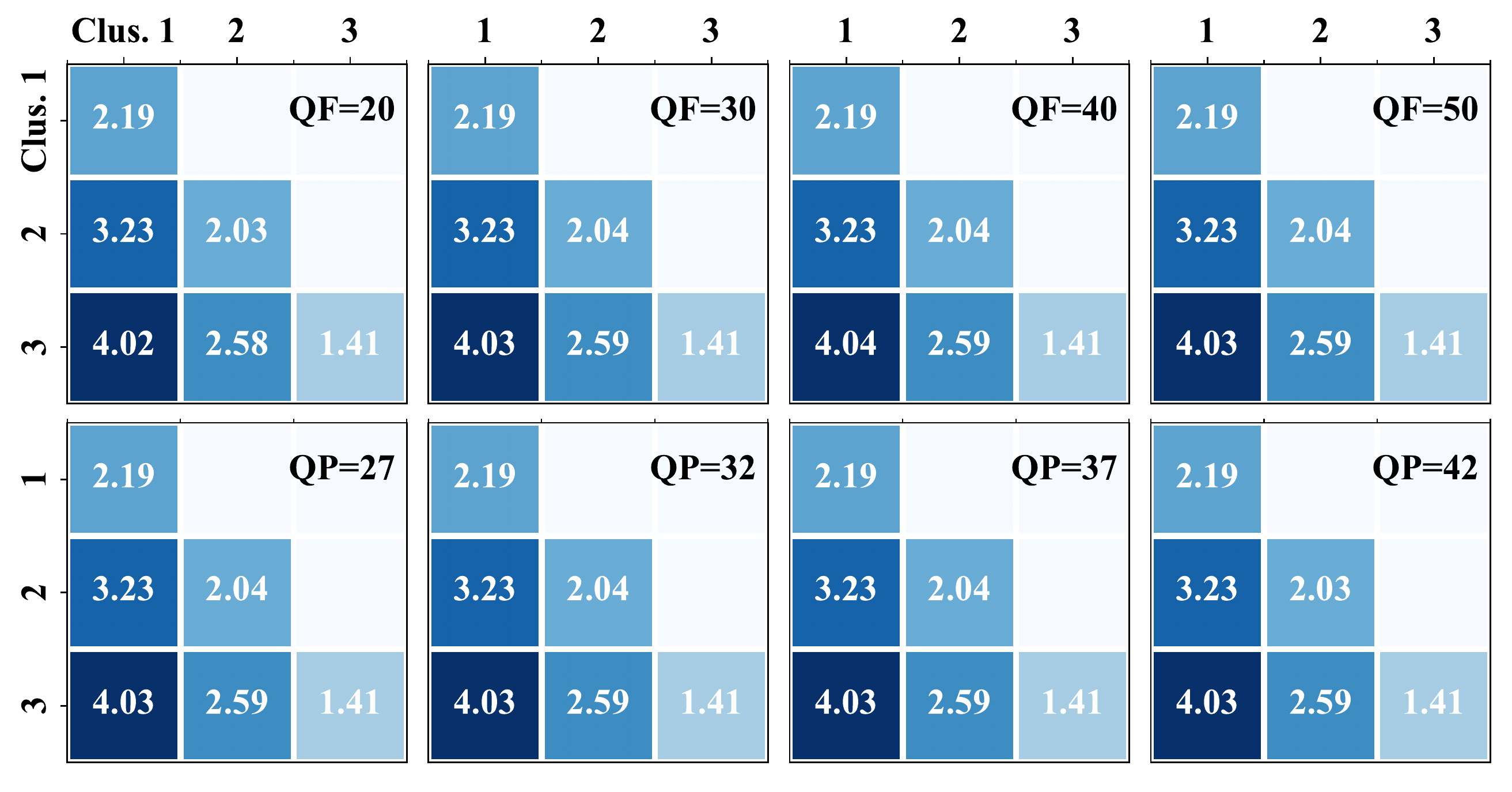}
  \caption{Texture difference between three clusters.
  The ``clus'' is the abbreviation of ``cluster''.
  The texture difference is measured by the TDIM value ($\times 10^3$).}
  \label{fig-observations-texture-difference}
\end{figure}

\textbf{Analysis:}
We cluster all patches into three clusters by the K-means clustering algorithm~\cite{lloydLeastSquaresQuantization1982,pedregosaScikitlearnMachineLearning2011} according to their defocus values.
Figure~\ref{fig-observations-defocus-cluster} shows that the patches in different clusters can differ significantly in quality, which accords with Observation 2.
Here, we further measure the average texture difference between patches in the same/different clusters.
Specifically, the texture difference of two patches is measured by the Frobenius norm of the difference between their Gram matrices of Y components, named the texture difference index measure (TDIM), which has been widely used in many texture-related works~\cite{gatysTextureSynthesisUsing2015,snelgroveHighresolutionMultiscaleNeural2017}.
Note that larger TDIM values indicate more diversity in the texture patterns between two patches.
As shown in Figure~\ref{fig-observations-texture-difference}, the TDIM values between patches in two different clusters are much larger than those between patches in the same cluster.
For example, for images compressed at QP $=37$, the average TDIM value between two patches in the first cluster is $2.19 \times 10^3$, significantly lower than that between two patches in clusters 1 and 3 (\ie, $4.03 \times 10^3$).
Therefore, for a compressed image, the texture patterns of patches with dissimilar defocus values are more diverse than those with similar defocus values.
Intuitively, the patches with dissimilar defocus values are blurred more diversely;
consequently, their texture patterns are also blurred more diversely than those with similar defocus values.
Finally, the analysis of Observation 3 is accomplished.

\section{Proposed approach}\label{sec-app}

\begin{figure*}[!t]
  \centering
  \includegraphics[width=1.\linewidth]{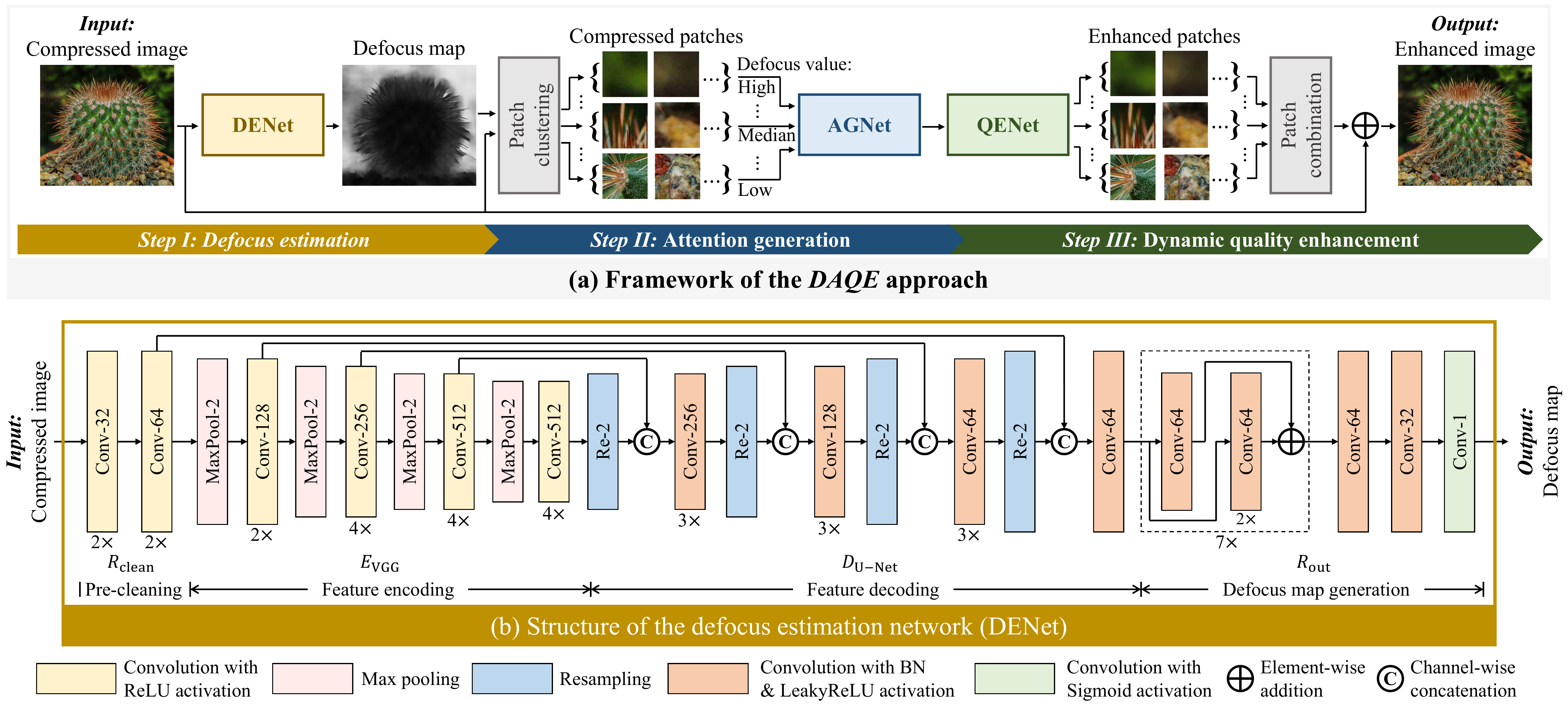}
  \caption{(a) Framework of the DAQE approach and (b) structure of the defocus estimation network (DENet).
    The notation ``Conv-$N$'' represents the convolution operator with $N$ output feature maps.
    The notation ``MaxPool/Re-$M$'' represents the max pooling/resampling operator with a factor of $M$.
    Notations ``BN'', ``ReLU'', and ``LeakyReLU'' represent the batch normalization~\cite{ioffeBatchNormalizationAccelerating2015}, rectified linear unit~\cite{nairRectifiedLinearUnits2010}, and leaky rectified linear unit, respectively.}
  \label{fig-daqe1}
\end{figure*}

In this section, we focus on our proposed DAQE approach for enhancing the quality of compressed images.
The DAQE approach aims to enhance the quality of regions with different defocus values.
Considering that these regions differ significantly in compression quality and texture patterns (as illustrated by Observations 2 and 3), we implement the DAQE approach by proposing an enhancement framework with three main steps as shown in Figure~\ref{fig-daqe1} (a), \ie, defocus estimation, attention generation, and dynamic quality enhancement.

Specifically, (1) the DAQE approach first estimates the defocus value for each image patch with a proposed defocus estimation network (DENet).
(2) Then, the DAQE approach divides patches into $N$ clusters according to their defocus values, and conducts cluster-specific texture extraction and quality enhancement.
To extract the texture pattern for each patch, the DAQE approach processes the input patch with a proposed attention generation network (AGNet).
AGNet consists of a convolution head and a transformer head to extract the texture pattern with local and global attention, respectively.
On the convolution head, local attention maps are generated to normalize the encoded feature of the input patch.
On the transformer head, the input patch is encoded and normalized by global attention to reference patches in the same cluster.
Next, the locally and globally normalized features are combined and serve as the texture pattern of the input patch.
(3) Finally, given the texture pattern of the input patch, the DAQE approach generates the enhanced patch with a proposed quality enhancement network (QENet).
QENet is equipped with a multilevel enhancement structure and works in a resource-efficient manner.
Specifically, clusters of patches with higher defocus values are simply enhanced by the former-level paths to save computational resources, while those with lower defocus values are further enhanced by the latter-level paths to achieve better quality.
All enhanced patches are spatially combined into the enhanced compressed image.
Given the above pipeline, the proposed DAQE approach can enhance the quality of compressed images effectively and efficiently by taking advantage of the inherent image defocus information.

\subsection{Defocus Estimation}\label{sec-sec-defocus-estimation}

In our DAQE approach, we design DENet to estimate a defocus map $\mathbf{M}$ for the input compressed image $\mathbf{I}_{\text{in}}$.
As shown in Figure~\ref{fig-daqe1} (b), DENet first adopts a series of residual blocks $R_{\text{clean}}$ to remove the severe compression artifacts of $\mathbf{I}_{\text{in}}$.
Hence, the precleaned feature $\mathbf{F}_{\text{clean}}$ is generated from $\mathbf{I}_{\text{in}}$, and is then encoded to be $\mathbf{F}_{\text{VGG}}$ by a VGG~\cite{simonyanVeryDeepConvolutional2015} encoder, denoted by $E_{\text{VGG}}$.
Here, $E_{\text{VGG}}$ is pretrained on ImageNet~\cite{dengImageNetLargescaleHierarchical2009} because pretraining on a large-scale dataset can facilitate the cross-domain learning of image defocus estimation (\ie, from the image domain to the defocus feature domain), as inspired by DMENet~\cite{leeDeepDefocusMap2019}.
Finally, a U-Net~\cite{ronnebergerUnetConvolutionalNetworks2015}-based decoder, denoted by $D_{\text{U-Net}}$, is adopted followed by a series of residual blocks $R_{\text{out}}$, for generating the defocus map $\mathbf{M}$ from $\mathbf{F}_{\text{VGG}}$.
Mathematically, we can obtain the defocus map $\mathbf{M}$ for the input compressed image $\mathbf{I}_{\text{in}}$ as follows:
\begin{equation}
  \mathbf{M} = R_{\text{out}} \left(
    D_{\text{U-Net}} \left(
      E_{\text{VGG}} \left(
        R_{\text{clean}} \left(
          \mathbf{I}_{\text{in}}
        \right)
      \right)
    \right)
  \right),
  \label{equ-defocus-estimation}
\end{equation}
where $\mathbf{M}$ and $\mathbf{I}_{\text{in}}$ have the same resolution.

Recall that the patchwise defocus value is the average defocus value for each patch.
Therefore, we can obtain the patchwise defocus value for each patch, by first dividing $\mathbf{M}$ into nonoverlapping $S \times S$ patches together with $\mathbf{I}_{\text{in}}$ and then calculating the average of the corresponding patch of $\mathbf{M}$.

\subsection{Defocus-Aware Attention Generation}\label{sec-sec-attention-generation}

\begin{figure*}[!t]
  \centering
  \includegraphics[width=1.\linewidth]{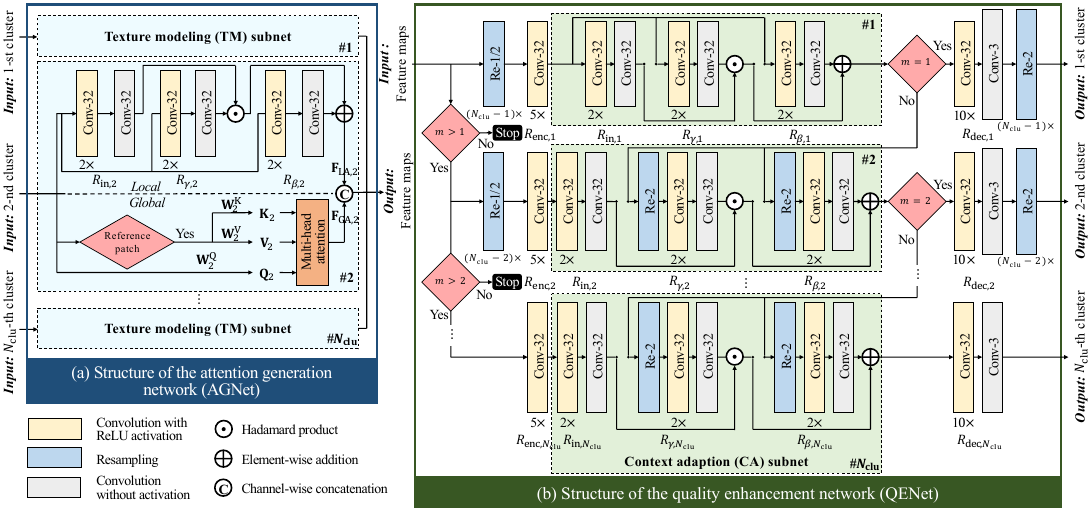}
  \caption{Structures of the (a) attention generation network (AGNet) and (b) quality enhancement network (QENet).
    The notation ``Conv-$N$'' represents the convolution operator with $N$ output feature maps.
    The notation ``Re-$M$'' represents the resampling operator with a factor of $M$.
    The notation ``ReLU'' represents the rectified linear unit~\cite{nairRectifiedLinearUnits2010}.}
  \label{fig-daqe2}
\end{figure*}

In our DAQE approach, we design AGNet to extract the texture pattern for an input patch.
The attention mechanism~\cite{vaswaniAttentionAllYou2017} has been widely used to extract texture patterns for image quality enhancement and other restoration tasks~\cite{zhangImageSuperresolutionUsing2018,wangRecoveringRealisticTexture2018,parkSemanticImageSynthesis2019,yangHiFaceGANFaceRenovation2020,zamirMultistageProgressiveImage2021}.
However, the various implementations of the attention mechanism in these works are conducted over the whole image, neglecting the texture diversity of different regions.
As shown in Figure~\ref{fig-daqe2} (a), AGNet mitigates this drawback by implementing the attention mechanism for regions with different defocus values separately, as those regions differ significantly in texture patterns, as discussed in Observation 3.
Specifically, AGNet first divides all patches into $N_{\text{clu}}$ clusters according to their defocus values.
Then, for each input patch, AGNet captures (1) local attention to the input patch and (2) global attention to reference patches in the same cluster.
Finally, AGNet encodes and normalizes the input patch with both the captured local and global attention, such that the texture pattern of the input patch can be obtained.

\textbf{Patch clustering.}
AGNet first divides all patches of an input image into $N_{\text{clu}}$ clusters according to their defocus values.
First, the defocus value centers of clusters are determined by the K-means algorithm~\cite{lloydLeastSquaresQuantization1982,pedregosaScikitlearnMachineLearning2011} over a large-scale dataset.
Then, every patch is assigned to its closest cluster, in terms of the sum-of-squares distances between its defocus value and the center defocus values.
In this way, AGNet can cluster the patches of the input image into different clusters with different defocus centers.
Notably, according to Observation 3, patches in different clusters possess diverse texture patterns.
This observation is utilized in the following attention generation.

\textbf{Local attention generation.}
As mentioned above, the attention mechanism has been widely used to extract texture patterns for image quality enhancement and other restoration tasks.
Inspired by the spatial adaptive normalization layer~\cite{parkSemanticImageSynthesis2019}, our AGNet includes a texture modeling subnet (TM subnet) for each cluster of patches, to extract the texture pattern for an input patch, as illustrated in Figure~\ref{fig-daqe2} (a).
We take the $m$-th TM subnet as an example, which processes the input patches in the $m$-th cluster, denoted by $\mathbf{P}_{m}$.
First, $\mathbf{P}_{m}$ is convolved by residual blocks $R_{\text{in},m}$ to obtain the encoded feature $R_{\text{in},m} \left(\mathbf{P}_{m}\right)$.
Then, $\mathbf{P}_{m}$ is convolved by residual blocks $R_{\gamma,m}$ and $R_{\beta,m}$, such that the corresponding attention maps $R_{\gamma,m} \left(\mathbf{P}_{m}\right)$ and $R_{\beta,m} \left(\mathbf{P}_{m}\right)$ can be produced.
Then, $R_{\text{in},m} \left(\mathbf{P}_{m}\right)$ is elementwise multiplied by $R_{\gamma,m} \left(\mathbf{P}_{m}\right)$ and then added by $R_{\beta,m} \left(\mathbf{P}_{m}\right)$ to generate the final output $\mathbf{F}_{\text{LA},m}$.
Mathematically, the above processes can be written as follows:
\begin{equation}
  \mathbf{F}_{\text{LA},m} = R_{\text{in},m} \left(
    \mathbf{P}_{m}
  \right) \odot R_{\gamma,m} \left(
    \mathbf{P}_{m}
  \right) + R_{\beta,m} \left(
    \mathbf{P}_{m}
  \right).
  \label{equ-local-attention}
\end{equation}
Note that patches in the same cluster share a TM subnet, \ie, with shared parameters, while those in different clusters are fed into different TM subnets, \ie, with different sets of parameters that are learned separately.
The reason is that the texture patterns of the patches in different clusters are more diverse than those in the same cluster.

\textbf{Global attention generation.}
Observation 3 reveals that the texture patterns of patches in the same cluster are more similar than those of patches in different clusters.
In light of this observation, AGNet includes a global branch in addition to the local branch for each TM subnet to take advantage of patches in the same cluster.
As depicted in Figure~\ref{fig-daqe2} (a), the global branch works through the following steps.
\begin{itemize}
  \item[1)]{$N_\text{ref}$ patches are proposed as the global reference patches~\cite{xiaVisionTransformerDeformable2022}, which generate the key/value pairs for the queries of all patches in their same cluster.
  AGNet first uniformly samples $\mathbf{I}_{\text{in}}$ to generate the initial reference patches, with a spatial sampling interval of $4S$ in both directions.
  Assume the height and width of $\mathbf{I}_{\text{in}}$ are $H$ and $W$, respectively.
  Then, we have $N_\text{ref}$ global reference patches, and $N_\text{ref}$ is equivalent to $H_s \times W_s$, where $H_s = \lfloor H / \left( 4S \right) \rfloor$ and $W_s = \lfloor W / \left( 4S \right) \rfloor$.}
  \item[2)]{AGNet adjusts the positions $\{\left( x_i, y_i \right)\}_{i=1}^{N_\text{ref}}$ of the initial reference patches by adding the position offsets $\{\left( \Delta x_i, \Delta y_i \right)\}_{i=1}^{N_\text{ref}}$.
  Here, $\{\left( \Delta x_i, \Delta y_i \right) \}_{i=1}^{N_\text{ref}}$ can be learned in the form of an offset map $\mathbf{F}_\text{offset}$ through an offset learning subnet as follows:
  \begin{equation}
    \mathbf{F}_\text{offset} = C_{1 \times 1} \left(
      L_{\text{GELU}} \left(
        C_{\text{DW}} \left(
          \mathbf{I}_{\text{in}}
        \right)
      \right)
    \right),
    \label{equ-reference-proposal}
  \end{equation}
  where $C_{1 \times 1}$, $L_{\text{GELU}}$, and $C_{\text{DW}}$ denote a convolution layer with a kernel size of $1 \times 1$, a GELU activation layer~\cite{hendrycksGaussianErrorLinear2016}, and a depthwise convolution layer, respectively.
  Note that $\mathbf{F}_\text{offset}$ is a map with size $H_s \times W_s$, in which each element denotes the offset pair $\left(\Delta x_i, \Delta y_i \right)$ for the $i$-th reference patch.}
  \item[3)]{AGNet performs differentiable sampling to obtain reference patches based on the initial reference patches $\{ \mathbf{P}^{\left( i \right)}_{\text{init}} \}_{i=1}^{N_\text{ref}}$ and the offset map $\mathbf{F}_\text{offset}$.
  Specifically, a reference patch $\mathbf{P}^{\left( i \right)}$ located at $\left( x_i + \Delta x_i, y_i + \Delta y_i \right)$ can be obtained by bilinearly sampling $\{ \mathbf{P}^{\left( i \right)}_{\text{init}} \}_{i=1}^{N_\text{ref}}$ as follows:
  \begin{equation}
    \small
    \mathbf{P}^{\left( i \right)} = \sum_{j=1}^{N_\text{ref}} g \left(x_i + \Delta x_i, x_j, W_s \right) g \left(y_i + \Delta y_i, y_j, H_s \right) \mathbf{P}^{\left( j \right)}_{\text{init}},
    \label{equ-bilinear-sampling}
  \end{equation}
  where $g(a, b, c) = 1 - | a - b | / c$.
  In this way, the acquisition of reference patches can be differentiable and trained in an end-to-end manner.}
  \item[4)]{Given the reference patches, AGNet computes the query of every input patch and the key/value pairs of the reference patches for each cluster as follows:
  \begin{align}
    \mathbf{Q}_{m} &= \widetilde{\mathbf{P}}_{m} \mathbf{W}_{m}^{\text{Q}},\\
    \mathbf{K}^{\left( i \right)}_{m} &= \widetilde{\mathbf{P}}_{m}^{\left( i \right)} \mathbf{W}_{m}^{\text{K},\left( i \right)}, i=1,2,\cdots,N_{\text{ref}}^{m},\\
    \mathbf{V}^{\left( i \right)}_{m} &= \widetilde{\mathbf{P}}_{m}^{\left( i \right)} \mathbf{W}_{m}^{\text{V},\left( i \right)}, i=1,2,\cdots,N_{\text{ref}}^{m}.
    \label{equ-reference-query}
  \end{align}
  In the above equations, $\widetilde{\mathbf{P}}_{m}$ and $\widetilde{\mathbf{P}}_{m}^{\left( i \right)}$ are the flattened $\mathbf{P}_{m}$ and the flattened $i$-th reference patch $\mathbf{P}_{m}^{\left( i \right)}$ in the $m$-th cluster, respectively;
  $\mathbf{W}_{m}^{\text{Q}}$, $\mathbf{W}_{m}^{\text{K},\left( i \right)}$ and $\mathbf{W}_{m}^{\text{V},\left( i \right)}$ are the projection matrices for the query, key, and value, respectively;
  $N_{\text{ref}}^{m}$ is the number of reference patches in the $m$-th cluster;
  $\mathbf{Q}_{m}$, $\mathbf{K}^{\left( i \right)}_{m}$ and $\mathbf{V}^{\left( i \right)}_{m}$ are the query of $\mathbf{P}_{m}$, key of $\mathbf{P}_{m}^{\left( i \right)}$, and value of $\mathbf{P}_{m}^{\left( i \right)}$, respectively.
  If $N_{\text{ref}}^{m}$ is equivalent to 0, we choose the reference patch in the neighboring cluster with the defocus value closest to the center of this cluster.}
  \item[5)]{AGNet performs multihead attention between $\mathbf{Q}_{m}$ and each $(\mathbf{K}_{m}^{(i)},\mathbf{V}_{m}^{(i)})$ pair.
  The attention output $\mathbf{Z}_{m}$ of each attention head can be formulated as,
  \begin{equation}
    \mathbf{Z}_{m} = \sum_{i=1}^{N_{\text{ref}}^m} \sigma \left(
      \mathbf{Q}_{m} {{\mathbf{K}}_{m}^{\left( i \right)}}^{\intercal} / \sqrt{d} + B
    \right) \mathbf{V}_{m}^{\left( i \right)}.
    \label{equ-reference-attention}
  \end{equation}
  In the above equation, $\sigma$ denotes the softmax function;
  $d$ is the dimension of each head;
  $B$ denotes the deformable relative position bias~\cite{xiaVisionTransformerDeformable2022}.}
\end{itemize}

Finally, $\mathbf{F}_{\text{GA},m}$ is generated by the multihead attention, which is then concatenated with $\mathbf{F}_{\text{LA},m}$, \ie, the output of the local branch.
This operation results in the output feature $\mathbf{F}_{\text{out}, m}$, which encodes the texture pattern for the input patch $\mathbf{P}_{m}$ via both local and global attention.
$\mathbf{F}_{\text{out}, m}$ is then sent to QENet as introduced in the next section.

\subsection{Defocus-Aware Dynamic Quality Enhancement}\label{sec-sec-quality-enhancement}

Given the estimated defocus value (Section~\ref{sec-sec-defocus-estimation}) and extracted texture pattern (Section~\ref{sec-sec-attention-generation}) for the input patch, our DAQE approach can finally conduct patchwise dynamic quality enhancement via the proposed QENet, as presented in the following.

\textbf{Dynamic structure with multilevel enhancement.}
The texture patterns in different clusters are more diverse than those in the same cluster, as revealed in Observation 3.
It is therefore effective to enhance different clusters of patches in a ``divide-and-conquer'' manner to finely restore the diverse texture patterns.
To this end, we equip QENet with a multilevel enhancement structure, which has $N_{\text{clu}}$ levels of enhancement paths as shown in Figure~\ref{fig-daqe2} (b).
In this dynamic structure, the input feature can be enhanced through different levels of paths, which are determined dynamically according to their defocus values.
These paths are not independent;
instead, they are connected progressively through context adaptation (CA) subnets.
Specifically, the feature of each path is adapted to the context information provided by the upper path to take advantage of the similarity in texture patterns between these two neighboring clusters.
QENet is resource-efficient in the following two aspects.
(1) Defocus-aware progressive enhancement.
The input features of patches with lower defocus values are enhanced via more levels of paths since they have inferior compression quality, as observed in Observation 2.
In the most sophisticated case, all levels of paths are traversed from top to bottom to achieve optimal enhancement performance.
Conversely, those with higher defocus values are enhanced by traversing only upper-level paths, so computational resources can be saved while maintaining high quality.
(2) Dynamic resolution of inference feature.
The input feature is downsampled by different factors at different levels before enhancement.
The enhanced image is finally upsampled to restore the resolution.
In this way, the upper-level enhancement is conducted over smaller inference features, thus consuming fewer computational resources.

\textbf{Quality enhancement at each level.}
Here, we take the enhancement of the input feature $\mathbf{F}_{\text{in},m}$ as an example, where $m$ is the cluster index.
Note that a cluster with a smaller $m$ has a higher center defocus value.
Let $n$ denote the level of the enhancement path.
Then, the paths of the top-$m$ levels are progressively traversed by $\mathbf{F}_{\text{in},m}$, \ie, level $n$ ranges from 1 to $m$.
Specifically, at the $n$-th level of enhancement, $\mathbf{F}_{\text{in},m}$ is first downsampled by a factor of 2 $(N_{\text{clu}} - n)$ times.
Then, the downsampled feature is encoded by residual blocks $R_{\text{enc},n}$ into $\mathbf{F}_{\text{enc},m}^{n}$ as follows:
\begin{equation}
  \mathbf{F}_{\text{enc},m}^{n} = R_{\text{enc},n} \bigl(
      \underbrace{\text{Dw} \bigl(
          \cdots \bigl(
            \text{Dw}}_{(N_{\text{clu}} - n)\,\text{times}} \bigl(
              \mathbf{F}_{\text{in},m}
            \bigr)
          \bigr)
        \cdots
    \bigr)
  \bigr),
  \label{equ-enhancement-encode}
\end{equation}
where Dw is a downsampling operator with a factor of $1/2$.
Subsequently, $\mathbf{F}_{\text{enc},m}^{n}$ is further processed by the CA subnet to adapt $\mathbf{F}_{\text{enc},m}^{n}$ to a context feature and generate an adapted feature $\mathbf{F}_{\text{ada},m}^{n}$.
For the top-level path, $\mathbf{F}_{\text{enc},m}^{1}$ serves as the context feature;
for other paths, the output of the CA subnet at the upper path $\mathbf{F}_{\text{ada},m}^{n-1}$ serves as the context feature.
The CA subnet first convolves $\mathbf{F}_{\text{enc},m}^{n}$ by a few residual blocks $\tilde{R}_{\text{in},n}$.
It then convolves the context feature by residual blocks $\tilde{R}_{\gamma,n}$ and $\tilde{R}_{\beta,n}$ to obtain the adaptation maps $\mathbf{F}_{\gamma,m}^{n}$ and $\mathbf{F}_{\beta,m}^{n}$, respectively.
The adapted feature $\mathbf{F}_{\text{ada},m}^{n}$ is produced by multiplying $\mathbf{F}_{\gamma,m}^{n}$ and then adding $\mathbf{F}_{\beta,m}^{n}$ to the convolved $\mathbf{F}_{\text{enc},m}^{n}$.
The above processes can be written as follows:
\begin{equation}
    \mathbf{F}_{\text{ada},m}^{n} = \tilde{R}_{\text{in},n} \bigl(
        \mathbf{F}_{\text{enc},m}^{n}
    \bigr) \odot \tilde{R}_{\gamma,n} \bigl(
        \mathbf{F}_{\text{ada},m}^{n-1}
    \bigr) + \tilde{R}_{\beta,n} \bigl(
        \mathbf{F}_{\text{ada},m}^{n-1}
    \bigr),
    \label{equ-enhancement-adaptation}
\end{equation}
where $\mathbf{F}_{\text{ada},m}^{0}$ refers to $\mathbf{F}_{\text{enc},m}^{1}$.
If the level index $n$ meets the cluster index $m$, $\mathbf{F}_{\text{ada},m}^{n}$ is further sent to the decoder, \ie, a set of residual blocks $R_{\text{dec},n}$, and then upsampled to generate the enhanced patch $\mathbf{P}_{\text{out},m}$ as follows:
\begin{equation}
  \mathbf{P}_{\text{out},m} = \underbrace{\text{Up} \bigl(
    \cdots \bigl(
      \text{Up}}_{(N_{\text{clu}} - n)\,\text{times}} \bigl(
        R_{\text{dec},n} \bigl(
          \mathbf{F}_{\text{ada},m}^{n}
        \bigr)
      \bigr)
    \bigr) \cdots
  \bigr),
  \label{equ-enhancement-decode}
\end{equation}
where Up is an upsampling operator with a factor of 2.
Finally, we can obtain the output enhanced image $\mathbf{I}_{\text{out}}$ for the input compressed image $\mathbf{I}_{\text{in}}$ by spatially combining all enhanced patches of all clusters.

\subsection{Loss Functions}\label{sec-sec-loss}

We train our DAQE model in a supervised manner.
Here, we discuss the loss functions for supervision, which are composed of a quality enhancement loss and a defocus estimation loss.

\textbf{Quality enhancement loss.}
Let $\mathcal{L}_{\text{en}}$ denote the quality enhancement loss.
Here, $\mathcal{L}_{\text{en}}$ is modeled by the Charbonnier loss function~\cite{charbonnierTwoDeterministicHalfquadratic1994} between the enhanced patch $\mathbf{P}_{\text{out}}$ and the corresponding raw patch $\hat{\mathbf{P}}$,
\begin{equation}
  \mathcal{L}_{\text{en}} = \sqrt{
    \Vert
      \mathbf{P}_{\text{out}} - \hat{\mathbf{P}}
    \Vert_2^2 + \epsilon^2,
  }
  \label{equ-loss-enhancement}
\end{equation}
where $\epsilon$ is a hyperparameter for numerical stability.
Then, AGNet and QENet are trained in an end-to-end manner by minimizing $\mathcal{L}_{\text{en}}$.

\textbf{Defocus estimation loss.}
We also take into account the defocus estimation loss $\mathcal{L}_{\text{de}}$ for training DENet.
Ideally, the ground truth defocus map for the compressed image is available for supervision.
Unfortunately, it is impossible to obtain the ground-truth defocus map for an image.
To solve this issue, we adopt the synthetic depth-of-field (SYNDOF) dataset~\cite{leeDeepDefocusMap2019} for training DENet.
The SYNDOF dataset contains 205 real defocused images $\mathbf{I}_{\text{real}}$ without ground-truth defocus maps.
It also contains 8,026 pairs of synthetic defocused image and defocus map $\{\mathbf{I}_{\text{syn}}, \hat{\mathbf{M}}\}$.
Note that $\mathbf{I}_{\text{syn}}$ are synthesized by the thin-lens model~\cite{potmesilLensApertureCamera1981} given $\hat{\mathbf{M}}$, as discussed in~\cite{leeDeepDefocusMap2019}.
Then, we compress $\mathbf{I}_{\text{real}}$ and $\mathbf{I}_{\text{syn}}$ into real compressed images $\mathbf{I}_{\text{real}}^c$ and synthetic compressed images  $\mathbf{I}_{\text{syn}}^c$, respectively.
Additional details about image compression are provided in Section~\ref{sec-sec-setup}.
Finally, we estimate the defocus maps $\mathbf{M}$ of $\mathbf{I}_{\text{syn}}^c$ by DENet and obtain a set of $\{\mathbf{I}_{\text{syn}}^c, \hat{\mathbf{M}}, \mathbf{M}\}$ for supervision.
Given the above training data of $\mathbf{I}_{\text{real}}^c$ and $\{\mathbf{I}_{\text{syn}}^c, \hat{\mathbf{M}}, \mathbf{M}\}$, we define the defocus estimation loss $\mathcal{L}_{\text{de}}$ as follows.
First, we minimize the pixelwise mean square error (MSE) between $\mathbf{M}$ and $\hat{\mathbf{M}}$,
\begin{equation}
  \mathcal{L}_{\text{pix}} = \Vert \mathbf{M} - \hat{\mathbf{M}} \Vert_2^2.
  \label{equ-loss-defocus-estimation-spatial}
\end{equation}
Then, we need to minimize the semantic distance between $\mathbf{M}$ and $\hat{\mathbf{M}}$, measured by the featurewise MSE,
\begin{equation}
  \mathcal{L}_{\text{feat}} = \Vert \phi (\mathbf{M}) - \phi (\hat{\mathbf{M}}) \Vert_2^2,
  \label{equ-loss-defocus-estimation-feat}
\end{equation}
where $\phi$ denotes the last convolution layer in the $l$-th block of a pretrained VGG-19 model~\cite{simonyanVeryDeepConvolutional2015}.
Next, we focus on reducing the domain gap between $\mathbf{I}_{\text{real}}^c$ and $\mathbf{I}_{\text{syn}}^c$ during defocus estimation through the following adversarial loss~\cite{goodfellowGenerativeAdversarialNets2014} between their feature maps:
\begin{equation}
  \mathcal{L}_{\text{adv}} = \alpha \cdot \log \bigl(
    \mathcal{D}( \psi( \mathbf{I}^c ))
  \bigr) + (1 - \alpha) \cdot \log \bigl(
    1 - \mathcal{D}( \psi( \mathbf{I}^c ))
  \bigr).
  \label{equ-loss-defocus-estimation-adversarial}
\end{equation}
In the above equation, $\mathbf{I}^c$ can be either $\mathbf{I}_{\text{real}}^c$ or $\mathbf{I}_{\text{syn}}^c$;
$\psi$ denotes the last upsampling layer of DENet;
$\mathcal{D}$ is a four-layer CNN-based discriminator;
$\alpha$ is a label, and it is equivalent to 0 when $\mathbf{I}^c = \mathbf{I}_{\text{real}}^c$ or is equivalent to 1 when $\mathbf{I}^c = \mathbf{I}_{\text{syn}}^c$.
Finally, the defocus estimation loss $\mathcal{L}_{\text{de}}$ is modeled as follows:
\begin{equation}
  \mathcal{L}_{\text{de}} = \mathcal{L}_{\text{pix}} + \lambda_{\text{feat}} \cdot \mathcal{L}_{\text{feat}} + \lambda_{\text{adv}} \cdot \mathcal{L}_{\text{adv}},
  \label{equ-loss-defocus-estimation}
\end{equation}
where $\lambda_{\text{feat}}$ and $\lambda_{\text{adv}}$ are the weight factors.
To obtain a converged discriminator, a discriminator loss $\mathcal{L}_{\mathcal{D}} = - \mathcal{L}_{\text{adv}}$ is set to supervise the training of $\mathcal{D}$.
Given the above loss functions, we can train DENet and the discriminator $\mathcal{D}$ by alternately minimizing $\mathcal{L}_{\text{de}}$ and $\mathcal{L}_{\mathcal{D}}$.

\section{Experiments}\label{sec-exp}

In this section, we present our experimental results to verify the performance of our proposed DAQE approach for the quality enhancement of compressed images.
Since BPG (HEVC-MSP)~\cite{bellardBetterPortableGraphics2018,sullivanOverviewHighEfficiency2012} and JPEG~\cite{wallaceJPEGStillPicture1992} are two widely used image compression codecs, our experiments focus on enhancing the quality of both BPG-compressed and JPEG-compressed images.

\subsection{Experimental Setup}\label{sec-sec-setup}

In this section, we present details about the datasets, hyperparameters, training strategy, and testing procedure of our DAQE approach.

\begin{table}[!t]
  \caption{Datasets adopted in this paper.
  The maximal image resolution (res.), image usage, and image indices of these datasets are indicated.}
  \label{tab-datasets}
  \centering
  \begin{tabular}{l | ccc}
    \toprule
    Dataset & Max res. & Usage & Indices \\
    \midrule
    DIV2K~\cite{agustssonNTIRE2017Challenge2017} & 2K & Training & 0001-0800 \\
    Kodak~\cite{kodakKodakLosslessTrue1999} & 768x512 & Testing & 0001-0025 \\
    DIV2K~\cite{agustssonNTIRE2017Challenge2017} & 2K & Testing & 0801-0900 \\
    Flickr2K~\cite{timofteNTIRE2017Challenge2017} & 2K & Testing & 2551-2650 \\
    RAISE~\cite{dang-nguyenRAISERawImages2015} & 4K & Testing & 8057-8156 \\
    \bottomrule
  \end{tabular}
\end{table}

\begin{table*}[!t]
  \caption{Quantitative comparison of our DAQE and compared approaches for BPG-compressed images.
  PSNR (dB) and SSIM are calculated with the BPG baseline as the anchor.
  Standard deviation values are presented in addition to the results.
  All results are calculated on the RGB channels.
  The PSNR and SSIM values are accurate to two and three decimal places, respectively.}
  \label{tab-efficacy-bpg}
  \centering
  \begin{tabular}{ r | c | cc | cc | cc | cc}
    \toprule
    \multirow{2}{*}{Approach} & \multirow{2}{*}{QP} & \multicolumn{2}{c|}{Kodak~\cite{kodakKodakLosslessTrue1999}} & \multicolumn{2}{c|}{DIV2K~\cite{agustssonNTIRE2017Challenge2017}} & \multicolumn{2}{c|}{Flickr2K~\cite{timofteNTIRE2017Challenge2017}} & \multicolumn{2}{c}{RAISE~\cite{dang-nguyenRAISERawImages2015}} \\
    & & PSNR & SSIM & PSNR & SSIM & PSNR & SSIM & PSNR & SSIM \\
    \midrule
    Baseline & \multirow{7}{*}{27} & 37.18$\pm$1.08 & 0.952$\pm$0.013 & 37.02$\pm$2.08 & 0.952$\pm$0.022 & 36.69$\pm$2.37 & 0.956$\pm$0.019 & 37.50$\pm$1.72 & 0.953$\pm$0.023 \\
    AR-CNN~\cite{dongCompressionArtifactsReduction2015} & & 37.52$\pm$1.06 & 0.954$\pm$0.013 & 37.63$\pm$1.99 & 0.956$\pm$0.021 & 37.23$\pm$2.13 & 0.960$\pm$0.017 & 37.97$\pm$1.68 & 0.958$\pm$0.021 \\
    DCAD~\cite{wangNovelDeepLearningbased2017} & & 37.75$\pm$1.08 & 0.956$\pm$0.012 & 37.90$\pm$1.98 & 0.958$\pm$0.021 & 37.48$\pm$2.08 & 0.962$\pm$0.017 & 38.22$\pm$1.67 & 0.959$\pm$0.021 \\
    DnCNN~\cite{zhangGaussianDenoiserResidual2017} & & 37.79$\pm$1.09 & 0.956$\pm$0.012 & 37.91$\pm$1.96 & 0.958$\pm$0.021 & 37.49$\pm$2.08 & 0.962$\pm$0.017 & 38.23$\pm$1.67 & 0.959$\pm$0.021 \\
    CBDNet~\cite{guoConvolutionalBlindDenoising2019} & & 37.96$\pm$1.07 & 0.957$\pm$0.012 & 38.14$\pm$1.95 & 0.959$\pm$0.021 & 37.75$\pm$2.01 & 0.963$\pm$0.017 & 38.29$\pm$1.69 & 0.960$\pm$0.021 \\
    RBQE~\cite{xingEarlyExitNot2020} & & 37.89$\pm$1.07 & 0.956$\pm$0.012 & 38.00$\pm$1.99 & 0.959$\pm$0.021 & 37.58$\pm$2.09 & 0.962$\pm$0.017 & 38.35$\pm$1.67 & 0.960$\pm$0.021 \\
    \rowcolor{mygray}
    \textbf{DAQE (Ours)} & & \textbf{38.27$\pm$1.11} & \textbf{0.958$\pm$0.012} & \textbf{38.45$\pm$1.94} & \textbf{0.960$\pm$0.021} & \textbf{38.03$\pm$1.97} & \textbf{0.964$\pm$0.017} & \textbf{38.66$\pm$1.68} & \textbf{0.962$\pm$0.021} \\
    \midrule
    Baseline & \multirow{7}{*}{32} & 33.97$\pm$1.35 & 0.915$\pm$0.019 & 34.15$\pm$2.19 & 0.921$\pm$0.035 & 33.76$\pm$2.44 & 0.928$\pm$0.025 & 34.44$\pm$1.96 & 0.922$\pm$0.029 \\
    AR-CNN~\cite{dongCompressionArtifactsReduction2015} & & 34.32$\pm$1.35 & 0.919$\pm$0.019 & 34.72$\pm$2.15 & 0.927$\pm$0.034 & 34.26$\pm$2.34 & 0.934$\pm$0.024 & 34.92$\pm$1.94 & 0.928$\pm$0.027 \\
    DCAD~\cite{wangNovelDeepLearningbased2017} & & 34.53$\pm$1.39 & 0.921$\pm$0.019 & 34.96$\pm$2.17 & 0.929$\pm$0.034 & 34.48$\pm$2.35 & 0.936$\pm$0.023 & 35.11$\pm$1.95 & 0.930$\pm$0.027 \\
    DnCNN~\cite{zhangGaussianDenoiserResidual2017} & & 34.57$\pm$1.39 & 0.921$\pm$0.019 & 34.98$\pm$2.16 & 0.929$\pm$0.034 & 34.50$\pm$2.35 & 0.936$\pm$0.023 & 35.14$\pm$1.95 & 0.931$\pm$0.027 \\
    CBDNet~\cite{guoConvolutionalBlindDenoising2019} & & 34.74$\pm$1.41 & 0.923$\pm$0.018 & 35.20$\pm$2.17 & 0.932$\pm$0.034 & 34.74$\pm$2.31 & 0.939$\pm$0.023 & 35.21$\pm$1.97 & 0.932$\pm$0.027 \\
    RBQE~\cite{xingEarlyExitNot2020} & & 34.66$\pm$1.39 & 0.922$\pm$0.019 & 35.06$\pm$2.18 & 0.931$\pm$0.034 & 34.58$\pm$2.35 & 0.938$\pm$0.023 & 35.23$\pm$1.95 & 0.932$\pm$0.027 \\
    \rowcolor{mygray}
    \textbf{DAQE (Ours)} & & \textbf{35.06$\pm$1.46} & \textbf{0.925$\pm$0.018} & \textbf{35.51$\pm$2.19} & \textbf{0.934$\pm$0.034} & \textbf{35.02$\pm$2.30} & \textbf{0.941$\pm$0.023} & \textbf{35.56$\pm$1.98} & \textbf{0.935$\pm$0.026} \\
    \midrule
    Baseline & \multirow{7}{*}{37} & 30.95$\pm$1.61 & 0.853$\pm$0.031 & 31.45$\pm$2.37 & 0.877$\pm$0.050 & 30.94$\pm$2.61 & 0.886$\pm$0.036 & 31.60$\pm$2.22 & 0.875$\pm$0.037 \\
    AR-CNN~\cite{dongCompressionArtifactsReduction2015} & & 31.27$\pm$1.63 & 0.858$\pm$0.031 & 31.97$\pm$2.36 & 0.885$\pm$0.049 & 31.39$\pm$2.61 & 0.893$\pm$0.035 & 32.02$\pm$2.22 & 0.882$\pm$0.036 \\
    DCAD~\cite{wangNovelDeepLearningbased2017} & & 31.44$\pm$1.66 & 0.861$\pm$0.032 & 32.17$\pm$2.39 & 0.888$\pm$0.049 & 31.58$\pm$2.63 & 0.896$\pm$0.035 & 32.19$\pm$2.25 & 0.885$\pm$0.036 \\
    DnCNN~\cite{zhangGaussianDenoiserResidual2017} & & 31.44$\pm$1.66 & 0.861$\pm$0.032 & 32.17$\pm$2.38 & 0.887$\pm$0.049 & 31.57$\pm$2.63 & 0.896$\pm$0.035 & 32.18$\pm$2.24 & 0.884$\pm$0.036 \\
    CBDNet~\cite{guoConvolutionalBlindDenoising2019} & & 31.65$\pm$1.71 & 0.864$\pm$0.032 & 32.40$\pm$2.40 & 0.891$\pm$0.048 & 31.81$\pm$2.65 & 0.899$\pm$0.035 & 32.28$\pm$2.27 & 0.887$\pm$0.036 \\
    RBQE~\cite{xingEarlyExitNot2020} & & 31.53$\pm$1.69 & 0.862$\pm$0.032 & 32.25$\pm$2.39 & 0.890$\pm$0.049 & 31.65$\pm$2.65 & 0.897$\pm$0.035 & 32.26$\pm$2.26 & 0.887$\pm$0.036 \\
    \rowcolor{mygray}
    \textbf{DAQE (Ours)} & & \textbf{31.95$\pm$1.75} & \textbf{0.868$\pm$0.032} & \textbf{32.69$\pm$2.44} & \textbf{0.895$\pm$0.049} & \textbf{32.07$\pm$2.69} & \textbf{0.903$\pm$0.035} & \textbf{32.61$\pm$2.32} & \textbf{0.892$\pm$0.036} \\
    \midrule
    Baseline & \multirow{7}{*}{42} & 28.36$\pm$1.87 & 0.766$\pm$0.055 & 29.04$\pm$2.61 & 0.817$\pm$0.065 & 28.35$\pm$2.88 & 0.822$\pm$0.054 & 29.07$\pm$2.53 & 0.809$\pm$0.056 \\
    AR-CNN~\cite{dongCompressionArtifactsReduction2015} & & 28.64$\pm$1.89 & 0.773$\pm$0.056 & 29.48$\pm$2.62 & 0.827$\pm$0.065 & 28.74$\pm$2.91 & 0.831$\pm$0.054 & 29.42$\pm$2.54 & 0.816$\pm$0.056 \\
    DCAD~\cite{wangNovelDeepLearningbased2017} & & 28.78$\pm$1.92 & 0.777$\pm$0.056 & 29.65$\pm$2.66 & 0.831$\pm$0.065 & 28.89$\pm$2.95 & 0.835$\pm$0.054 & 29.54$\pm$2.58 & 0.820$\pm$0.056 \\
    DnCNN~\cite{zhangGaussianDenoiserResidual2017} & & 28.80$\pm$1.93 & 0.777$\pm$0.056 & 29.68$\pm$2.67 & 0.832$\pm$0.064 & 28.91$\pm$2.96 & 0.836$\pm$0.054 & 29.57$\pm$2.59 & 0.821$\pm$0.056 \\
    CBDNet~\cite{guoConvolutionalBlindDenoising2019} & & 28.96$\pm$1.97 & 0.781$\pm$0.057 & 29.87$\pm$2.71 & 0.836$\pm$0.064 & 29.10$\pm$3.01 & 0.840$\pm$0.054 & 29.60$\pm$2.61 & 0.822$\pm$0.056 \\
    RBQE~\cite{xingEarlyExitNot2020} & & 28.85$\pm$1.95 & 0.778$\pm$0.057 & 29.71$\pm$2.67 & 0.833$\pm$0.065 & 28.94$\pm$2.99 & 0.837$\pm$0.055 & 29.60$\pm$2.61 & 0.822$\pm$0.057 \\
    \rowcolor{mygray}
    \textbf{DAQE (Ours)} & & \textbf{29.19$\pm$2.00} & \textbf{0.786$\pm$0.058} & \textbf{30.08$\pm$2.76} & \textbf{0.840$\pm$0.064} & \textbf{29.28$\pm$3.06} & \textbf{0.844$\pm$0.055} & \textbf{29.88$\pm$2.68} & \textbf{0.829$\pm$0.057} \\
    \bottomrule
  \end{tabular}
\end{table*}

\textbf{Datasets.}
Recent works have adopted some large-scale image datasets, such as BSDS500~\cite{arbelaezContourDetectionHierarchical2011} and ImageNet~\cite{dengImageNetLargescaleHierarchical2009}, for image denoising, segmentation, and other image tasks.
However, the images from these datasets contain unknown artifacts, since they are collected under unknown conditions and compressed by unknown codecs and settings.
To obtain ``clean'' images without significant artifacts, we adopt several high-quality image datasets for evaluation, as illustrated in Table~\ref{tab-datasets}.
Specifically, we adopt 800 images of the DIV2K dataset~\cite{agustssonNTIRE2017Challenge2017} as the training set.
In addition, we adopt all 25 images of the Kodak dataset~\cite{kodakKodakLosslessTrue1999}, 100 images of the DIV2K dataset, 100 images of the Flickr2K dataset~\cite{timofteNTIRE2017Challenge2017}, and 100 images of the RAISE dataset~\cite{dang-nguyenRAISERawImages2015} as the test set.
We compress all images using the BPG~\cite{bellardBetterPortableGraphics2018} and JPEG codecs~\cite{wallaceJPEGStillPicture1992}.
We adopt four compression settings for each codec, \ie, the quantization parameter (QP) is 27/32/37/42 in BPG and the quality factor (QF) is 20/30/40/50 in JPEG.
These settings are widely used for other quality enhancement works~\cite{wangNovelDeepLearningbased2017,yangMultiframeQualityEnhancement2018,guanMFQENewApproach2021,xingEarlyExitNot2020}.

\textbf{Hyperparameters, training and testing.}
In our DAQE approach, $S$, $N_{\text{clu}}$, and $d$ are set to 128, 3, and 32, respectively.
The number of attention heads is set to 3.
All convolution operators have a kernel size of 3, a stride of 1, and padding of 1.
To cluster the input patches, we adopt the K-means clustering algorithm~\cite{lloydLeastSquaresQuantization1982,pedregosaScikitlearnMachineLearning2011} over the DIV2K training set.
For the loss functions, we set $\epsilon$, $l$, $\lambda_{\text{feat}}$, and $\lambda_{\text{adv}}$ to $10^{-6}$, 4, $10^{-4}$, and $10^{-3}$, respectively.
During the training process, the Adam~\cite{kingmaAdamMethodStochastic2017} optimizer is applied with an initial learning rate of $10^{-4}$.
The cosine annealing schedule~\cite{loshchilovSGDRStochasticGradient2017} is applied to decrease the learning rate automatically.
The training batch size is set to 64.
A workstation with one CPU (Intel Xeon Platinum 8163 CPU @ 2.50GHz) and four GPUs (Tesla V100-SXM2-16GB) is used for training and testing.
We first train DENet on the training set of SYNDOF.
After the convergence of DENet, we freeze the parameters of DENet and train the subsequent AGNet and QENet jointly on the DIV2K training set until convergence.

\subsection{Evaluation}\label{sec-sec-eval}

In this section, we evaluate the performance of our DAQE approach for the quality enhancement of compressed images.
We compare our approach with several widely used approaches including AR-CNN~\cite{dongCompressionArtifactsReduction2015}, DCAD~\cite{wangNovelDeepLearningbased2017}, DnCNN~\cite{zhangGaussianDenoiserResidual2017}, CBDNet~\cite{guoConvolutionalBlindDenoising2019} and RBQE~\cite{xingEarlyExitNot2020}.
Among them, CBDNet and RBQE were originally used for blind restoration.
For fair comparisons, we retrain them in a nonblind manner, \ie, train one model for each compression configuration.
In addition, all compared approaches are retrained on our training set.\footnote{Codes of all approaches are available at \url{https://github.com/RyanXingQL/PowerQE}.}

\begin{table*}[!t]
  \caption{Quantitative comparison of our DAQE and compared approaches for JPEG-compressed images.
  PSNR (dB) and SSIM are calculated with the JPEG baseline as the anchor.
  Standard deviation values are presented in addition to the results.
  All results are calculated on the RGB channels.
  The PSNR and SSIM values are accurate to two and three decimal places, respectively.}
  \label{tab-efficacy-jpeg}
  \centering
  \begin{tabular}{ r | c | cc | cc | cc | cc}
    \toprule
    \multirow{2}{*}{Approach} & \multirow{2}{*}{QF} & \multicolumn{2}{c|}{Kodak~\cite{kodakKodakLosslessTrue1999}} & \multicolumn{2}{c|}{DIV2K~\cite{agustssonNTIRE2017Challenge2017}} & \multicolumn{2}{c|}{Flickr2K~\cite{timofteNTIRE2017Challenge2017}} & \multicolumn{2}{c}{RAISE~\cite{dang-nguyenRAISERawImages2015}} \\
    & & PSNR & SSIM & PSNR & SSIM & PSNR & SSIM & PSNR & SSIM \\
    \midrule
    Baseline & \multirow{7}{*}{20} & 29.04$\pm$1.99 & 0.828$\pm$0.029 & 29.59$\pm$2.75 & 0.851$\pm$0.050 & 28.83$\pm$3.15 & 0.855$\pm$0.040 & 29.70$\pm$2.64 & 0.852$\pm$0.037 \\
    AR-CNN~\cite{dongCompressionArtifactsReduction2015} & & 30.31$\pm$2.15 & 0.855$\pm$0.033 & 31.03$\pm$2.87 & 0.881$\pm$0.051 & 30.15$\pm$3.37 & 0.883$\pm$0.041 & 30.97$\pm$2.78 & 0.876$\pm$0.036 \\
    DCAD~\cite{wangNovelDeepLearningbased2017} & & 30.63$\pm$2.21 & 0.862$\pm$0.034 & 31.37$\pm$2.93 & 0.888$\pm$0.051 & 30.47$\pm$3.45 & 0.890$\pm$0.040 & 31.25$\pm$2.85 & 0.883$\pm$0.035 \\
    DnCNN~\cite{zhangGaussianDenoiserResidual2017} & & 30.71$\pm$2.23 & 0.863$\pm$0.034 & 31.45$\pm$2.93 & 0.888$\pm$0.051 & 30.54$\pm$3.47 & 0.891$\pm$0.040 & 31.32$\pm$2.86 & 0.884$\pm$0.035 \\
    CBDNet~\cite{guoConvolutionalBlindDenoising2019} & & 30.93$\pm$2.28 & 0.867$\pm$0.034 & 31.74$\pm$3.02 & 0.893$\pm$0.050 & 30.81$\pm$3.54 & 0.895$\pm$0.039 & 31.26$\pm$2.93 & 0.885$\pm$0.035 \\
    RBQE~\cite{xingEarlyExitNot2020} & & 30.79$\pm$2.29 & 0.863$\pm$0.034 & 31.60$\pm$3.04 & 0.891$\pm$0.051 & 30.64$\pm$3.57 & 0.893$\pm$0.040 & 31.44$\pm$2.98 & 0.888$\pm$0.035 \\
    \rowcolor{mygray}
    \textbf{DAQE (Ours)} & & \textbf{31.28$\pm$2.35} & \textbf{0.871$\pm$0.035} & \textbf{32.02$\pm$3.06} & \textbf{0.897$\pm$0.050} & \textbf{31.06$\pm$3.62} & \textbf{0.899$\pm$0.039} & \textbf{31.82$\pm$3.04} & \textbf{0.893$\pm$0.035} \\
    \midrule
    Baseline & \multirow{7}{*}{30} & 30.38$\pm$2.04 & 0.864$\pm$0.023 & 30.91$\pm$2.88 & 0.880$\pm$0.045 & 30.20$\pm$3.35 & 0.886$\pm$0.035 & 31.05$\pm$2.70 & 0.882$\pm$0.033 \\
    AR-CNN~\cite{dongCompressionArtifactsReduction2015} & & 31.65$\pm$2.19 & 0.885$\pm$0.025 & 32.37$\pm$2.98 & 0.904$\pm$0.045 & 31.51$\pm$3.48 & 0.908$\pm$0.034 & 32.31$\pm$2.80 & 0.902$\pm$0.030 \\
    DCAD~\cite{wangNovelDeepLearningbased2017} & & 32.00$\pm$2.25 & 0.892$\pm$0.026 & 32.73$\pm$3.03 & 0.910$\pm$0.045 & 31.85$\pm$3.54 & 0.914$\pm$0.033 & 32.62$\pm$2.87 & 0.908$\pm$0.029 \\
    DnCNN~\cite{zhangGaussianDenoiserResidual2017} & & 32.07$\pm$2.26 & 0.893$\pm$0.026 & 32.81$\pm$3.04 & 0.910$\pm$0.045 & 31.91$\pm$3.55 & 0.915$\pm$0.032 & 32.69$\pm$2.86 & 0.909$\pm$0.029 \\
    CBDNet~\cite{guoConvolutionalBlindDenoising2019} & & 32.26$\pm$2.31 & 0.896$\pm$0.026 & 33.06$\pm$3.05 & 0.914$\pm$0.044 & 32.18$\pm$3.58 & 0.918$\pm$0.032 & 32.59$\pm$2.92 & 0.909$\pm$0.029 \\
    RBQE~\cite{xingEarlyExitNot2020} & & 32.11$\pm$2.31 & 0.893$\pm$0.026 & 32.89$\pm$3.06 & 0.912$\pm$0.045 & 31.98$\pm$3.61 & 0.916$\pm$0.033 & 32.77$\pm$2.97 & 0.911$\pm$0.029 \\
    \rowcolor{mygray}
    \textbf{DAQE (Ours)} & & \textbf{32.64$\pm$2.38} & \textbf{0.900$\pm$0.026} & \textbf{33.39$\pm$3.09} & \textbf{0.918$\pm$0.044} & \textbf{32.47$\pm$3.63} & \textbf{0.922$\pm$0.032} & \textbf{33.18$\pm$3.00} & \textbf{0.916$\pm$0.029} \\
    \midrule
    Baseline & \multirow{7}{*}{40} & 31.30$\pm$2.06 & 0.885$\pm$0.020 & 31.79$\pm$2.93 & 0.896$\pm$0.041 & 31.13$\pm$3.46 & 0.903$\pm$0.032 & 31.97$\pm$2.75 & 0.900$\pm$0.029 \\
    AR-CNN~\cite{dongCompressionArtifactsReduction2015} & & 32.57$\pm$2.19 & 0.903$\pm$0.021 & 33.24$\pm$2.99 & 0.917$\pm$0.041 & 32.42$\pm$3.50 & 0.922$\pm$0.029 & 33.22$\pm$2.81 & 0.917$\pm$0.027 \\
    DCAD~\cite{wangNovelDeepLearningbased2017} & & 32.93$\pm$2.25 & 0.908$\pm$0.021 & 33.62$\pm$3.05 & 0.922$\pm$0.041 & 32.77$\pm$3.55 & 0.927$\pm$0.028 & 33.55$\pm$2.87 & 0.922$\pm$0.026 \\
    DnCNN~\cite{zhangGaussianDenoiserResidual2017} & & 33.00$\pm$2.26 & 0.909$\pm$0.021 & 33.69$\pm$3.04 & 0.923$\pm$0.041 & 32.83$\pm$3.54 & 0.927$\pm$0.028 & 33.61$\pm$2.86 & 0.923$\pm$0.026 \\
    CBDNet~\cite{guoConvolutionalBlindDenoising2019} & & 33.19$\pm$2.30 & 0.912$\pm$0.021 & 33.94$\pm$3.03 & 0.926$\pm$0.040 & 33.10$\pm$3.55 & 0.930$\pm$0.027 & 33.50$\pm$2.89 & 0.923$\pm$0.026 \\
    RBQE~\cite{xingEarlyExitNot2020} & & 33.03$\pm$2.29 & 0.909$\pm$0.021 & 33.75$\pm$3.05 & 0.924$\pm$0.041 & 32.88$\pm$3.58 & 0.928$\pm$0.029 & 33.67$\pm$2.93 & 0.924$\pm$0.026 \\
    \rowcolor{mygray}
    \textbf{DAQE (Ours)} & & \textbf{33.56$\pm$2.37} & \textbf{0.915$\pm$0.022} & \textbf{34.29$\pm$3.07} & \textbf{0.929$\pm$0.040} & \textbf{33.40$\pm$3.58} & \textbf{0.933$\pm$0.027} & \textbf{34.10$\pm$2.95} & \textbf{0.929$\pm$0.026} \\
    \midrule
    Baseline & \multirow{7}{*}{50} & 32.05$\pm$2.04 & 0.899$\pm$0.017 & 32.49$\pm$2.94 & 0.909$\pm$0.038 & 31.89$\pm$3.62 & 0.915$\pm$0.030 & 32.70$\pm$2.74 & 0.912$\pm$0.027 \\
    AR-CNN~\cite{dongCompressionArtifactsReduction2015} & & 33.29$\pm$2.16 & 0.915$\pm$0.018 & 33.93$\pm$2.98 & 0.927$\pm$0.038 & 33.13$\pm$3.51 & 0.931$\pm$0.027 & 33.93$\pm$2.78 & 0.927$\pm$0.025 \\
    DCAD~\cite{wangNovelDeepLearningbased2017} & & 33.64$\pm$2.21 & 0.920$\pm$0.018 & 34.29$\pm$3.03 & 0.931$\pm$0.038 & 33.47$\pm$3.53 & 0.936$\pm$0.025 & 34.25$\pm$2.82 & 0.932$\pm$0.024 \\
    DnCNN~\cite{zhangGaussianDenoiserResidual2017} & & 33.68$\pm$2.21 & 0.920$\pm$0.018 & 34.33$\pm$3.01 & 0.931$\pm$0.037 & 33.50$\pm$3.52 & 0.936$\pm$0.025 & 34.28$\pm$2.80 & 0.931$\pm$0.024 \\
    CBDNet~\cite{guoConvolutionalBlindDenoising2019} & & 33.91$\pm$2.27 & 0.923$\pm$0.018 & 34.61$\pm$3.01 & 0.934$\pm$0.037 & 33.81$\pm$3.51 & 0.939$\pm$0.024 & 34.22$\pm$2.84 & 0.932$\pm$0.024 \\
    RBQE~\cite{xingEarlyExitNot2020} & & 33.72$\pm$2.25 & 0.920$\pm$0.018 & 34.40$\pm$3.02 & 0.932$\pm$0.037 & 33.56$\pm$3.56 & 0.937$\pm$0.025 & 34.35$\pm$2.87 & 0.933$\pm$0.024 \\
    \rowcolor{mygray}
    \textbf{DAQE (Ours)} & & \textbf{34.32$\pm$2.33} & \textbf{0.927$\pm$0.018} & \textbf{35.01$\pm$3.03} & \textbf{0.938$\pm$0.037} & \textbf{34.15$\pm$3.53} & \textbf{0.942$\pm$0.024} & \textbf{34.86$\pm$2.88} & \textbf{0.938$\pm$0.024} \\
    \bottomrule
  \end{tabular}
\end{table*}

\textbf{Quantitative performance.}
To evaluate the efficacy of our DAQE approach, we measure PSNR and SSIM for different approaches on both BPG-compressed and JPEG-compressed images from four different datasets.
Table~\ref{tab-efficacy-bpg} presents the results on BPG-compressed images. 
As shown in Table~\ref{tab-efficacy-bpg}, the average PSNR of the DAQE approach on the DIV2K dataset is 32.69 dB at QP $=37$, which is 1.24 dB higher than that of the BPG baseline and 0.29 dB higher than that of the second-best approach.
In addition, the average SSIM is 0.895, which is 0.018 higher than the BPG baseline and 0.004 higher than that of the second-best approach.
Similar results can be found for the other three datasets and other QP settings.
For the JPEG-compressed images, Table~\ref{tab-efficacy-jpeg} shows that the average PSNR of the DAQE approach on the DIV2K dataset is 34.29 dB at QF $=40$, which is 2.50 dB higher than that of the JPEG baseline, and 0.35 dB higher than that of the second-best approach.
In addition, the average SSIM is 0.929, which is 0.033 higher than that of the JPEG baseline and 0.003 higher than that of the second-best approach.
Similar results can be found for the other three datasets and other QF settings.
In summary, the DAQE approach achieves state-of-the-art performance on all four datasets for both BPG-compressed and JPEG-compressed images.

\begin{figure*}[!t]
  \centering
  \includegraphics[width=1.\linewidth]{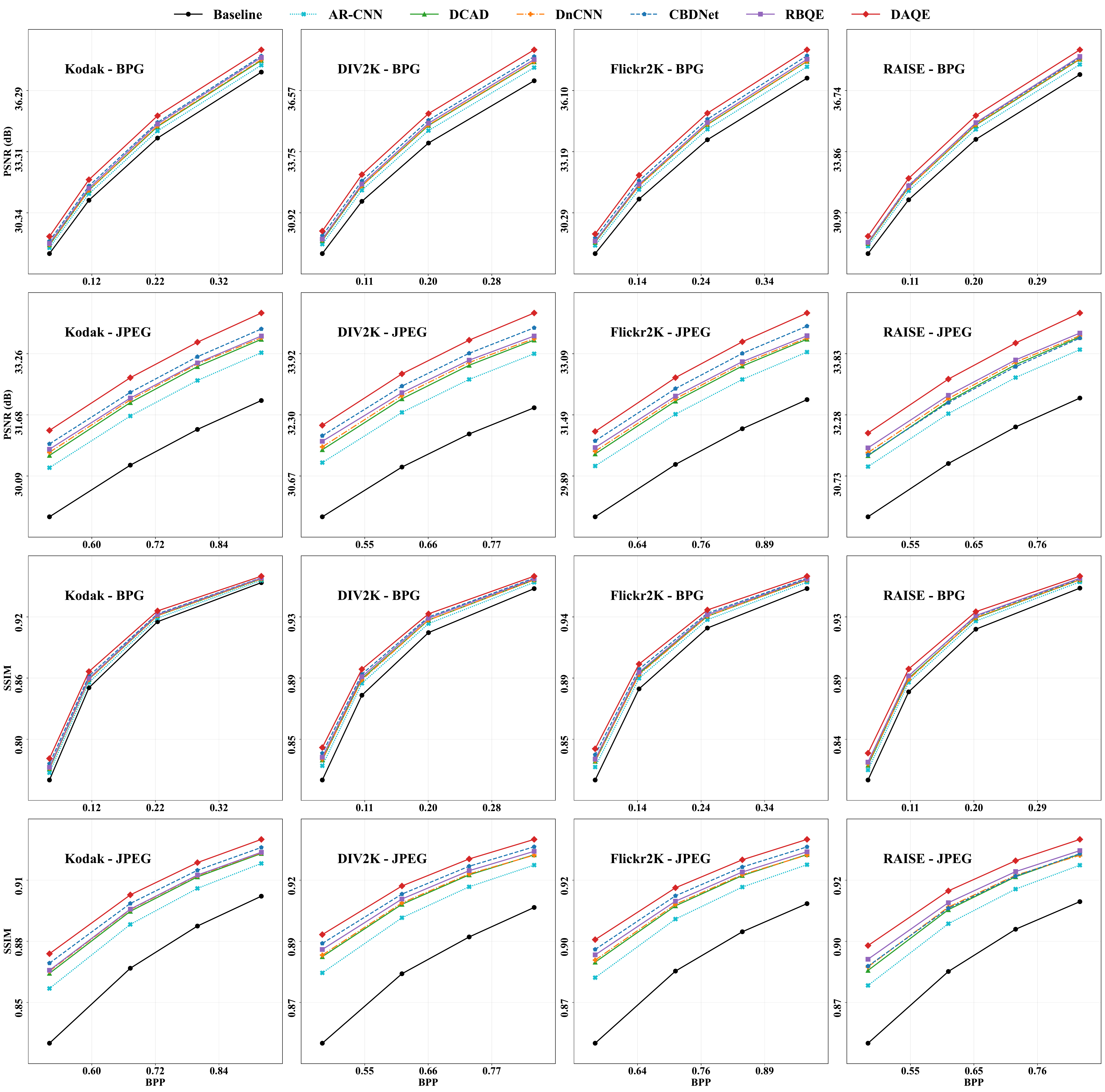}
  \caption{Rate-distortion curves of our DAQE and compared approaches.
  The rate is measured by the bits per pixel (BPP).
  The distortion is measured by PSNR (dB) and SSIM.}
  \label{fig-rate-distortion}
\end{figure*}

\begin{table*}[!t]
  \caption{Rate-distortion performance of our DAQE and compared approaches.
  The rate-distortion performance is measured by the BD-rate reduction (\%) with the BPG/JPEG baseline as the anchor.
  Standard deviations are presented in addition to the results.
  The rate is measured by the bits per pixel (BPP).
  The distortion is measured by PSNR (dB) and SSIM.}
  \label{tab-rate-distortion}
  \centering
  \begin{tabular}{ r | cc | cc | cc | cc}
    \toprule
    \multirow{2}{*}{Approach} & \multicolumn{2}{c|}{Kodak~\cite{kodakKodakLosslessTrue1999}} & \multicolumn{2}{c|}{DIV2K~\cite{agustssonNTIRE2017Challenge2017}} & \multicolumn{2}{c|}{Flickr2K~\cite{timofteNTIRE2017Challenge2017}} & \multicolumn{2}{c}{RAISE~\cite{dang-nguyenRAISERawImages2015}} \\
    & BPG & JPEG & BPG & JPEG & BPG & JPEG & BPG & JPEG \\
    \midrule
    \multicolumn{9}{c}{BPP-PSNR} \\
    \midrule
    AR-CNN~\cite{dongCompressionArtifactsReduction2015} & -7.06$\pm$1.93 & -21.22$\pm$2.93 & -11.14$\pm$5.64 & -23.48$\pm$6.11 & -9.26$\pm$4.80 & -21.26$\pm$5.26 & -9.45$\pm$5.46 & -20.78$\pm$5.73 \\
    DCAD~\cite{wangNovelDeepLearningbased2017} & -10.85$\pm$3.01 & -25.88$\pm$3.91 & -15.20$\pm$6.75 & -28.11$\pm$7.26 & -12.84$\pm$5.88 & -25.66$\pm$6.31 & -12.91$\pm$6.45 & -24.86$\pm$6.57 \\
    DnCNN~\cite{zhangGaussianDenoiserResidual2017} & -11.24$\pm$2.98 & -26.79$\pm$3.98 & -15.47$\pm$6.94 & -29.08$\pm$7.77 & -12.99$\pm$5.94 & -26.46$\pm$6.42 & -13.19$\pm$6.39 & -25.81$\pm$6.70 \\
    CBDNet~\cite{guoConvolutionalBlindDenoising2019} & -14.79$\pm$3.95 & -29.64$\pm$4.53 & -19.38$\pm$7.81 & -32.53$\pm$7.79 & -16.88$\pm$7.05 & -29.92$\pm$7.42 & -14.65$\pm$6.67 & -24.77$\pm$6.82 \\
    RBQE~\cite{xingEarlyExitNot2020} & -12.91$\pm$3.12 & -27.72$\pm$4.34 & -16.82$\pm$7.43 & -30.92$\pm$10.02 & -14.28$\pm$6.10 & -27.47$\pm$6.67 & -14.66$\pm$6.95 & -27.16$\pm$7.28 \\
    \rowcolor{mygray}
    \textbf{DAQE (Ours)} & \textbf{-20.17$\pm$5.09} & \textbf{-34.22$\pm$5.65} & \textbf{-24.05$\pm$9.10} & \textbf{-35.17$\pm$10.02} & \textbf{-21.08$\pm$8.16} & \textbf{-33.01$\pm$8.21} & \textbf{-20.66$\pm$7.99} & \textbf{-32.19$\pm$8.00} \\
    \midrule
    \multicolumn{9}{c}{BPP-SSIM} \\
    \midrule
    AR-CNN~\cite{dongCompressionArtifactsReduction2015} & -5.18$\pm$2.46 & -17.48$\pm$5.67& -10.40$\pm$8.13 & -22.91$\pm$10.18& -9.03$\pm$6.33 & -20.57$\pm$7.64& -9.09$\pm$8.42 & -17.90$\pm$7.60 \\
    DCAD~\cite{wangNovelDeepLearningbased2017} & -8.36$\pm$3.76 & -22.36$\pm$6.62& -13.55$\pm$14.07 & -28.35$\pm$12.03& -12.68$\pm$8.10 & -25.59$\pm$8.72& -12.57$\pm$10.03 & -22.74$\pm$8.53 \\
    DnCNN~\cite{zhangGaussianDenoiserResidual2017} & -8.25$\pm$3.71 & -22.93$\pm$6.65& -14.36$\pm$8.95 & -29.28$\pm$13.23& -12.23$\pm$7.73 & -26.19$\pm$9.14& -12.05$\pm$9.44 & -25.03$\pm$15.20 \\
    CBDNet~\cite{guoConvolutionalBlindDenoising2019} & -11.17$\pm$4.81 & -25.46$\pm$7.54& -16.45$\pm$21.19 & -32.58$\pm$13.57& -16.15$\pm$9.81 & -29.88$\pm$10.40& -14.72$\pm$12.33 & -23.71$\pm$10.94 \\
    RBQE~\cite{xingEarlyExitNot2020} & -9.26$\pm$4.45 & -23.25$\pm$7.48& -16.29$\pm$10.70 & -31.17$\pm$13.81& -13.99$\pm$9.58 & -28.08$\pm$11.17& -14.57$\pm$13.31 & -26.44$\pm$13.78 \\
    \rowcolor{mygray}
    \textbf{DAQE (Ours)} & \textbf{-14.89$\pm$6.25} & \textbf{-28.65$\pm$8.74} & \textbf{-22.14$\pm$11.77} & \textbf{-34.32$\pm$12.19} & \textbf{-19.74$\pm$11.17} & \textbf{-32.50$\pm$10.81} & \textbf{-19.89$\pm$14.51} & \textbf{-30.11$\pm$13.64} \\
    \bottomrule
  \end{tabular}
\end{table*}

\textbf{Rate-distortion performance.}
We further evaluate the rate-distortion performance of our DAQE approach in Figure~\ref{fig-rate-distortion} and Table~\ref{tab-rate-distortion}.
Figure~\ref{fig-rate-distortion} shows the rate-distortion curves of different approaches on the four datasets.
As shown in the figure,  the rate-distortion curves of our DAQE approach are higher than those of other approaches, indicating the superior rate-distortion performance of our approach.
Then, we quantify the rate-distortion performance by evaluating the reduction in Bjontegaard-rate (BD-rate)~\cite{bjontegaardCalculationAveragePSNR2001}.
The results are presented in Table~\ref{tab-rate-distortion}.
As shown, for BPG-compressed images, the BD-rate reductions of our DAQE approach on the DIV2K dataset are on average 24.05\% and 22.14\% with the distortion measured by PSNR and SSIM, respectively, while those of the second-best approach are only 19.38\% and 16.45\% on average.
Similar results can be observed for the other three datasets and JPEG-compressed images.
In summary, our DAQE approach significantly surpasses state-of-the-art rate-distortion performance.

\begin{figure*}[!t]
  \centering
  \includegraphics[width=.8\linewidth]{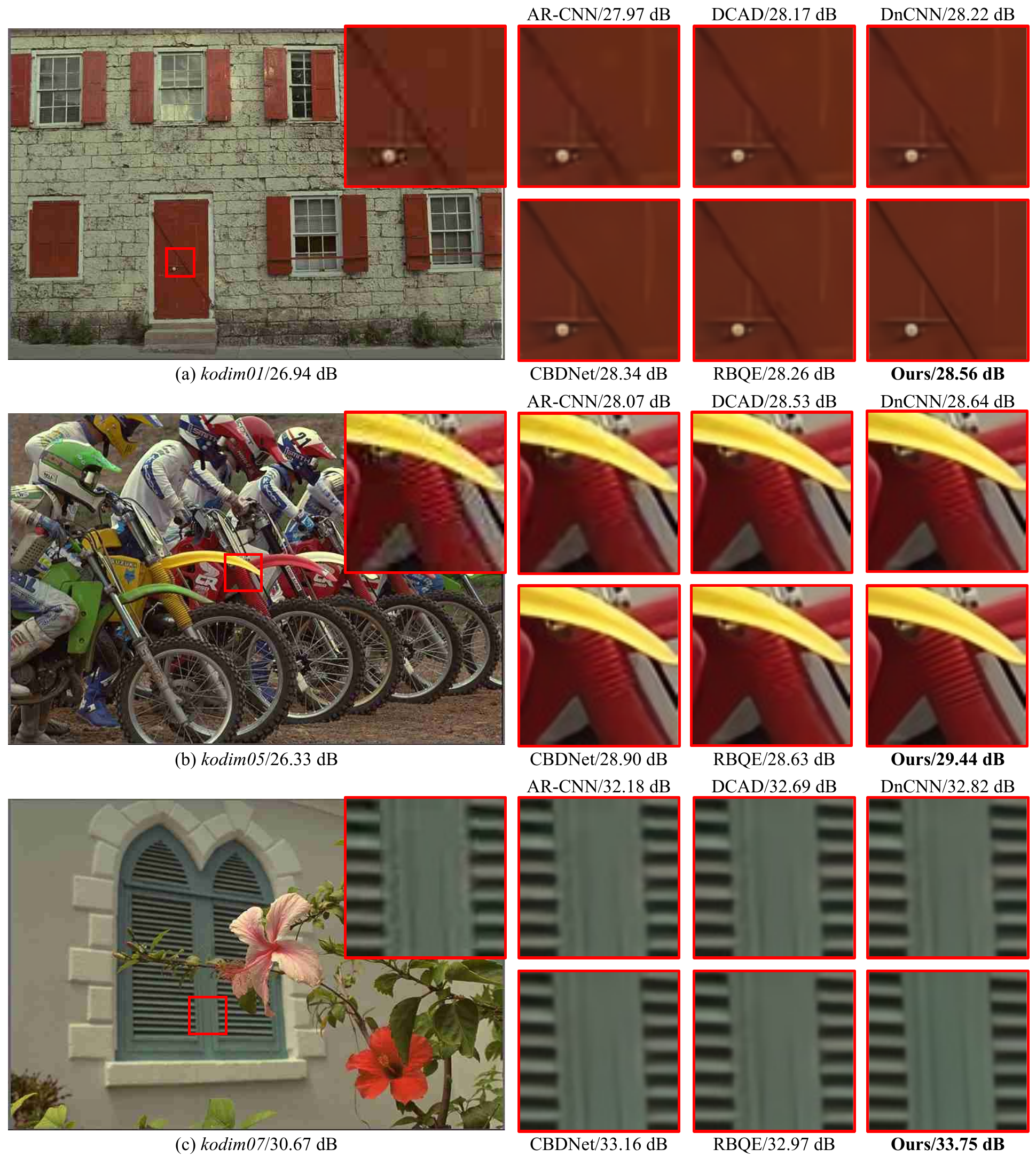}
  \caption{Qualitative comparison of our DAQE and compared approaches.}
  \label{fig-subjective-quality}
\end{figure*}

\textbf{Qualitative performance.}
Figure~\ref{fig-subjective-quality} compares the visual results of our DAQE and the compared approaches.
Specifically, the DAQE approach successfully restores the edge details of the door, motorbike, and window in Figure~\ref{fig-subjective-quality} (a)-(c), respectively.
In contrast, these details cannot be well restored by the other approaches.
In addition, the DAQE approach suppresses the compression artifacts around these edges, while those artifacts are hardly reduced by the other approaches.
To summarize, the DAQE approach outperforms the compared approaches qualitatively, especially in restoring details and suppressing compression artifacts.

\begin{figure}[!t]
  \centering
  \includegraphics[width=.95\linewidth]{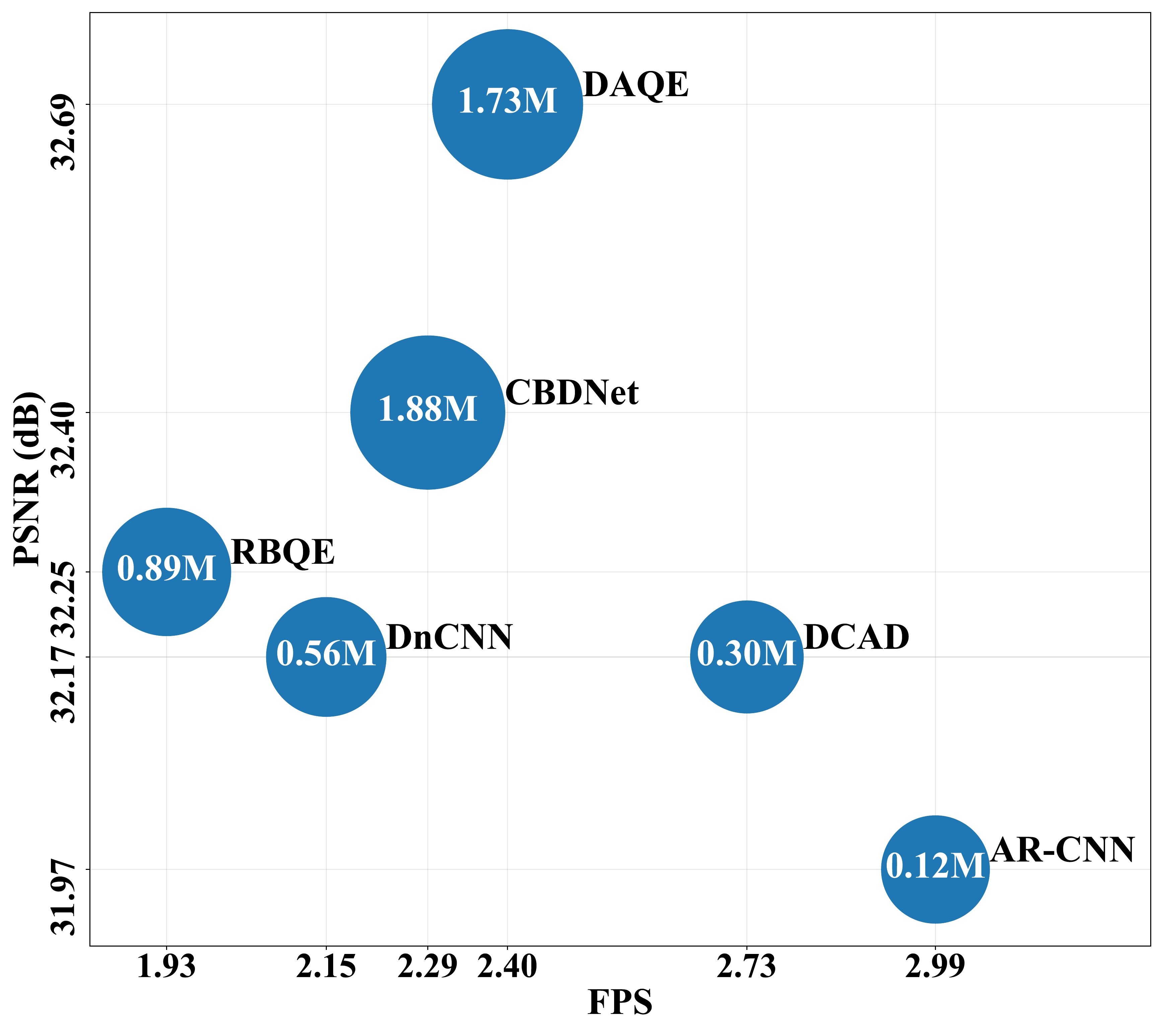}
  \caption{Efficiency of our DAQE and compared approaches over the DIV2K test set compressed by BPG at QP $=37$.
  The number of parameters is marked at the center of each circle.
  A larger circle radius indicates a larger number of parameters.}
  \label{fig-efficiency}
\end{figure}

\textbf{Efficiency.}
We measure the efficiency of our DAQE and other compared approaches from two aspects: the time complexity in terms of the frames per second (FPS) and the space complexity in terms of the number of parameters.
As shown in Figure~\ref{fig-efficiency}, the DAQE approach outperforms the second-best CBDNet by 0.29 dB in PSNR with 8.67\% fewer parameters and 4.80\% higher FPS.
Some approaches, such as AR-CNN and DCAD, have fewer parameters and higher FPS than the DAQE approach and CBDNet.
However, their PSNR performance is at least 0.44 dB lower than that of the DAQE approach.
In summary, our DAQE approach achieves a good balance between efficiency and enhancement performance.

\textbf{Defocus estimation.}
To evaluate the defocus estimation performance of DENet for compressed images, we compare DENet with the state-of-the-art DMENet~\cite{leeDeepDefocusMap2019} on the CUHK dataset compressed by BPG at QP$=$37.
In addition to the officially released model of DMENet, we also retrain DMENet on the official training set but with compression, named DMENet-Comp.
Finally, we measure the accuracy of defocus estimation by each model.
The average accuracy scores of DENet, DMENet, and DMENet-Comp are 81.73\%, 70.96\%, and 76.30\%, respectively.
In other words, the compression artifacts degrade the defocus estimation accuracy of the compared DMENet by 5.34\%.
In addition, the proposed DENet outperforms the retrained DMENet by 5.49\% in accuracy on the test set.
These experimental results further demonstrate the necessity of proposing DENet to estimate the defocus map for compressed images.

\subsection{Ablation Study}\label{sec-sec-ablation}

\begin{table}[!t]
  \caption{Ablation results of our DAQE approach in terms of PSNR (dB).}
  \label{tab-ablation}
  \centering
  \begin{tabular}{l | cccc}
    \toprule
    Component & \textbf{DAQE} & (A) & (B) & (C) \\
    \midrule
    Local attention of AGNet & \cmark & \cmark & \cmark & \xmark \\
    Global attention of AGNet & \cmark & \cmark & \xmark & \xmark \\
    CA subnet of QENet & \cmark & \xmark & \xmark & \xmark \\
    \midrule
    PSNR (dB) & \textbf{32.69} & 32.59 & 32.54 & 32.51 \\
    \bottomrule
  \end{tabular}
\end{table}

\textbf{Network components.}
Some important components are proposed in our DAQE network.
First, a local attention module is designed in AGNet to extract the texture pattern for each input patch.
In addition, AGNet is equipped with a global attention module for extracting the texture pattern of the input patch by referring to all patches in the same cluster.
Finally, the CA subnet is proposed to effectively connect each level of QENet.
To validate the effectiveness of these network components, we gradually ablate each component to generate three different networks denoted by (A) to (C), as presented in Table~\ref{tab-ablation}.
Then, we retrain and test all these networks on the DIV2K dataset compressed by BPG at QP $=37$.
As shown in Table~\ref{tab-ablation}, ablating the CA subnets degrades PSNR by 0.10 dB.
Further ablations of the global attention and local attention lead to 0.05 and 0.03 dB degradation in PSNR, respectively.
Thus, these network components have a positive influence on the enhancement performance of the DAQE approach.

\begin{figure}[!t]
  \centering
  \includegraphics[width=1.\linewidth]{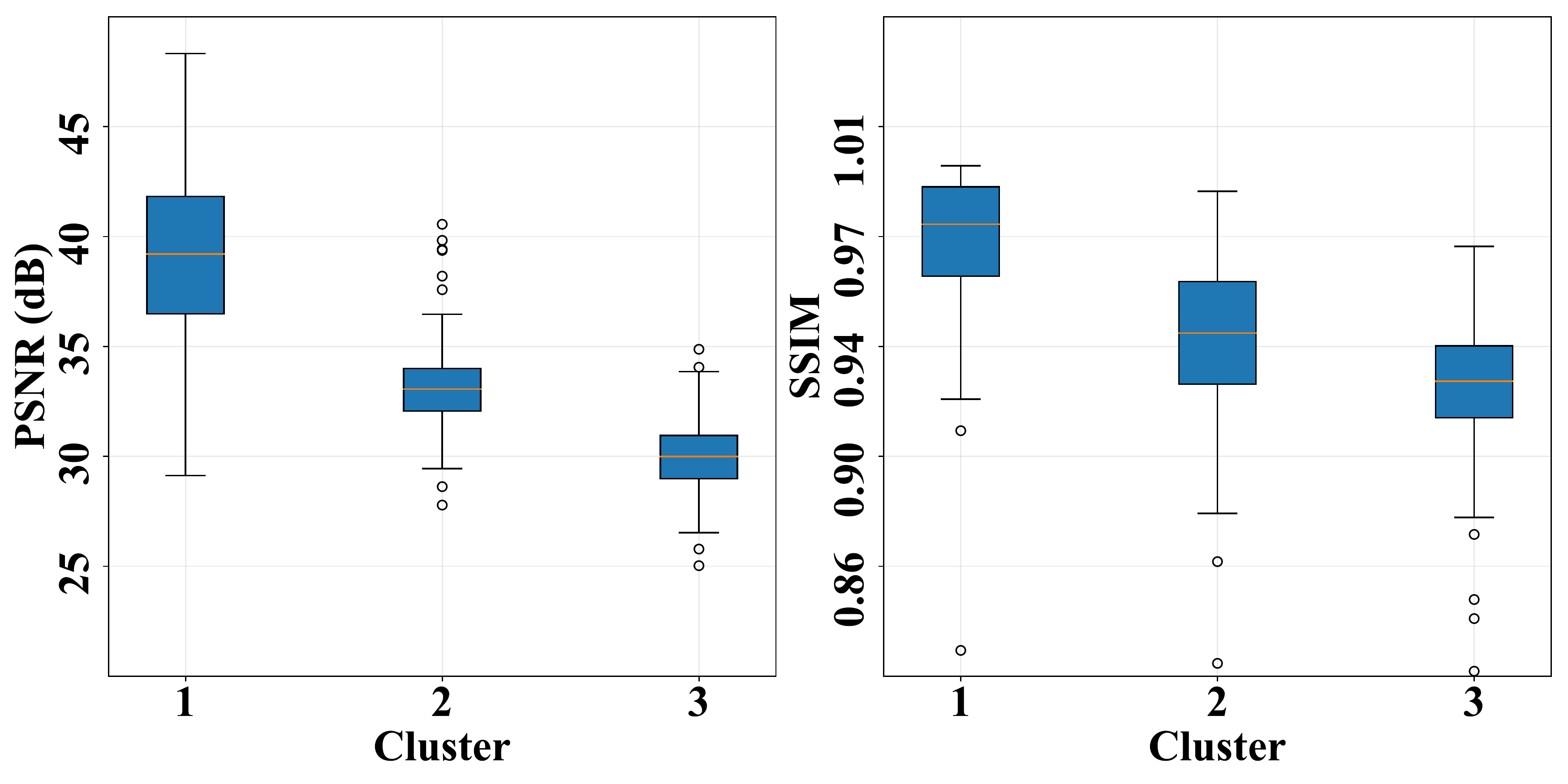}
  \caption{Statistics of PSNR and SSIM values for patches in different clusters.
  Patches are from the DIV2K test set compressed by BPG at QP $=37$.}
  \label{fig-patch-classification}
\end{figure}

\textbf{Defocus-based patch classification.}
During the process of defocus-based patch classification, we classify the patches into three different clusters by DENet, to prepare for subsequent dynamic quality enhancement.
Figure~\ref{fig-patch-classification} shows the statistics of the patches after defocus estimation and clustering.
Patches from different clusters significantly differ in compression quality in terms of both the PSNR and SSIM values.
Specifically, the median PSNR values of the patches in the three clusters are 39.21, 33.06, and 30.00 dB.
The median SSIM values of the patches in the three clusters are 0.98, 0.94, and 0.92.
Large quality gaps between patches in different clusters bring great benefits to cluster-specific quality enhancement.
Notably, the classification of compressed patches in our DAQE approach does not rely on raw patches, making it practical in real-world applications.

We also measure the upper bound for our DAQE approach (\ie, DAQE-Upper) by computing the compression quality of patches in terms of PSNR and then clustering patches according to their quality.
Note that we retrain DAQE-Upper on the training set.
Experimental results show that DAQE-Upper achieves an average PSNR of 32.84 dB, which further improves the performance of DAQE (\ie, 32.69 dB) by 0.15 dB.
This experiment provides the upper bound for DAQE and demonstrates the effectiveness of enhancing patches with different quality in a divide-and-conquer manner.
More importantly, by reasoning about image defocus, DAQE can efficiently cluster patches with different quality without requiring raw patches.
In summary, the above experiment demonstrates the effectiveness of using defocus for quality enhancement in the aspect of clustering patches with different quality.

\textbf{Defocus-based attention.}
To evaluate the effectiveness of defocus-based attention, we design a defocus-blind quality enhancement approach, called DAQE-Blind, with the following modifications to DAQE.
(1) First, all reference patches are used for the global attention module, since there is only one cluster and all reference patches are adopted for this cluster.
(2) Second, all patches are forced to exit at a fixed level of QENet, because DAQE-Blind cannot manage the dynamic inference of DAQE without knowing the defocus information.
For a fair comparison, we exit all patches at the first level of QENet for DAQE-Blind and DAQE.
We then train these two approaches on our training set and evaluate their performance.

We measure the PSNR-FPS performance of these two approaches.
DAQE-Blind achieves an average PSNR of 32.47 dB, which is slightly worse than DAQE (\ie, 32.56 dB).
In other words, the enhanced PSNR degrades by 8.11\% for DAQE-Blind compared with DAQE, \ie, 1.02 vs. 1.11 dB.
More importantly, the FPS results of DAQE-Blind and DAQE are 1.85 and 2.81, respectively, indicating a speed degradation of 34.16\% for DAQE-Blind over DAQE.
The reason for this degradation is that DAQE uses only a few reference patches with similar quality and texture, but all reference patches are presented to DAQE-Blind, making it more difficult for DAQE-Blind to find relevance from a large number of patches and then learn from those patches.
In summary, it is necessary to exploit the defocus characteristic of image patches in our approach in terms of both effectiveness and efficiency.

Finally, we measure the correlation between defocus differences and attention ranks for each input patch of DAQE-Blind.
Specifically, the defocus differences are measured between the input patch and all reference patches;
the attention ranks are obtained by referring to the attention values of the reference patches.
The experimental result shows that the PCC value can reach 0.74 on average.
Therefore, the attended patches by DAQE-Blind are similar to the input patch in terms of defocus.
In other words, a small number of patches with similar defocus values can serve as effective references for quality enhancement to find regionwise relevance.
The above experiments show the effectiveness of using defocus for quality enhancement in the aspect of finding regionwise relevance.

\begin{figure}[!t]
  \centering
  \includegraphics[width=.95\linewidth]{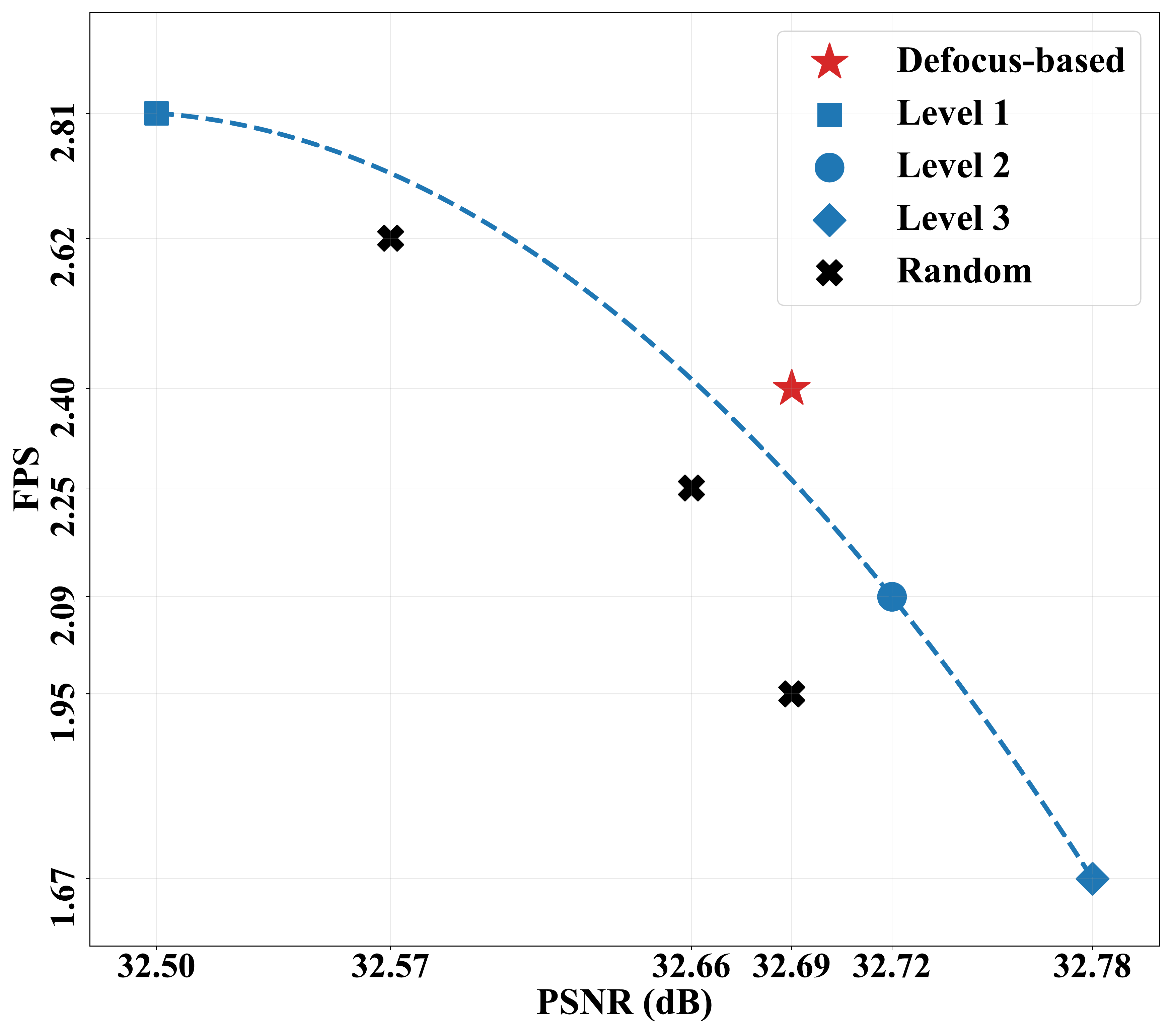}
  \caption{PSNR-FPS performance of different enhancement strategies over the DIV2K test set compressed by BPG at QP $=37$.}
  \label{fig-envelope}
\end{figure}

\textbf{Defocus-based dynamic enhancement.}
To evaluate the efficacy of the defocus-based dynamic enhancement of the DAQE approach, we design the following experiments.
Specifically, instead of the defocus-based dynamic enhancement, we compulsively exit all patches at the first level of QENet without considering their defocus values.
All patches are treated in a single cluster and all reference patches are used for the global attention module.
The PSNR-FPS result is denoted by the blue square in Figure~\ref{fig-envelope}.
Similarly, all patches exit at the second and the third levels of QENet separately, generating two PSNR-FPS results also shown in Figure~\ref{fig-envelope}.
Finally, each patch randomly exits with three different random seeds, generating three PSNR-FPS results, as shown in Figure~\ref{fig-envelope}.
As shown, compared with other strategies, our defocus-based dynamic enhancement (denoted by the red star) achieves a superior tradeoff between enhancement quality and speed.

\textbf{Frequency-based clustering.}
The quality and texture pattern of compressed patches are also related to their frequency content.
Therefore, we equip the proposed image restoration architecture with frequency detection and frequency-based patch clustering and then verify its performance.
The resulting approach is denoted by DAQE-Freq.
Specifically, DAQE-Freq computes the wavelet energy~\cite{wangMWGANPerceptualQuality2022} of image patches and then clusters the patches according to their energy.
Here, the wavelet energy is computed as the summed squares of the wavelet coefficients of the high-frequency subbands, \ie, LH, HL, and HH.
We then train DAQE-Freq on our training set.

We measure the PSNR performance of DAQE-Freq.
The average PSNR over the test set is 32.52 dB, which is 0.17 dB lower than that of DAQE (\ie, 32.69 dB).
We also find that the average PCC value between the detected frequency and PSNR values is 0.70, which is worse than that between the estimated defocus and PSNR values (\ie, 0.78).
The reason is that the frequency detection is performed on the compressed patches and is thus affected by compression artifacts.
To demonstrate this fact, we feed DAQE-Freq with the ``clean'' wavelet energy of raw patches and then retrain DAQE-Freq.
The PCC value increases from 0.70 to 0.76, and the PSNR increases from 32.52 dB to 32.62 dB.
Note that the raw patches cannot be obtained in practice during enhancement.
In summary, simply replacing the defocus estimation with frequency detection degrades the performance of our DAQE approach for enhancing the quality of compressed images.

\section{Conclusion}

In this paper, we proposed the defocus-aware quality enhancement (DAQE) approach.
Our DAQE approach considers the regionwise defocus difference of compressed images, thus differing from the traditional quality enhancement approaches in two aspects.
(1) The DAQE approach employs fewer computational resources to enhance the quality of significantly defocused regions and more resources to enhance the quality of other regions.
(2) The DAQE approach learns to separately enhance diverse texture patterns for regions with different defocus values, such that texture-specific enhancement can be managed.
To achieve these goals, the DAQE approach first estimates the defocus value for each image region with the proposed DENet.
Next, patches are classified into different clusters according to their defocus values and then sent to AGNet and QENet to accomplish cluster-specific texture extraction and dynamic quality enhancement.
Finally, extensive experiments validated that our DAQE approach can significantly improve the quality of compressed images in a resource-efficient manner and is superior to existing state-of-the-art approaches.

We propose two research directions for future work.
(1) Our work considers PSNR and SSIM as the metrics for compression quality to be enhanced.
Future work could embrace other perceptual quality metrics to improve the QoE of compressed images since the image defocus also correlates with the perceptual quality of compressed images.
(2) Our work focuses on the quality enhancement of compressed images.
Future work may extend the scope of defocus-aware approaches to other image enhancement and restoration tasks, \eg, image denoising and deblurring, because image defocus is inherent in the physics of image formation and can be utilized by more low-level vision tasks.


%




\ifCLASSOPTIONcaptionsoff
  \newpage
\fi



\bibliographystyle{IEEEtran}
\typeout{}
\bibliography{IEEEabrv,refs}
\end{document}